\documentclass[twocolumn,iop]{emulateapj}
\usepackage{natbib, color, amsmath, ulem, epstopdf}
\usepackage[caption=false]{subfig}

\begin{document}

\title{ SEEDS Adaptive Optics Imaging of the Asymmetric Transition Disk Oph IRS 48 in Scattered Light \thanks{Based on data collected at Subaru Telescope, which is operated by the National Astronomical Observatory of Japan.} }

\author{Katherine B. Follette\altaffilmark{1},
Carol A. Grady\altaffilmark{2},
Jeremy R. Swearingen\altaffilmark{3,4}, 
Michael L. Sitko\altaffilmark{3,4,5}, 
Elizabeth H. Champney\altaffilmark{3,4},
Nienke van der Marel\altaffilmark{6}, 
Michihiro Takami\altaffilmark{7}, 
Marc  J Kuchner\altaffilmark{8}, 
Laird M. Close\altaffilmark{1}, 
Takayuki Muto\altaffilmark{9}, 
Satoshi Mayama\altaffilmark{10},
Michael W. McElwain\altaffilmark{8}, 
Misato Fukagawa\altaffilmark{11}, 
Koen Maaskant\altaffilmark{6}, 
Michiel Min\altaffilmark{6}, 
Ray W. Russell\altaffilmark{12,4}, 
Tomoyuki Kudo\altaffilmark{13}, 
Nobuhiko Kusakabe\altaffilmark{14},
Jun Hashimoto\altaffilmark{15}, 
Lyu Abe\altaffilmark{16},
Eiji Akiyama\altaffilmark{14},
Wolfgang Brandner\altaffilmark{17},
Timothy D. Brandt\altaffilmark{18},
Joseph Carson\altaffilmark{19}, Thayne Currie \altaffilmark{13},
Sebastian E. Egner\altaffilmark{13},
Markus Feldt\altaffilmark{17},
Miwa Goto\altaffilmark{20},
Olivier Guyon\altaffilmark{13},
Yutaka Hayano\altaffilmark{13},
Masahiko Hayashi\altaffilmark{14},
Saeko Hayashi\altaffilmark{13},
Thomas Henning\altaffilmark{17},
Klaus Hodapp\altaffilmark{21},
Miki Ishii\altaffilmark{14}, 
Masanori Iye\altaffilmark{14},
Markus Janson\altaffilmark{22},
Ryo Kandori\altaffilmark{14},
Gillian R. Knapp\altaffilmark{18},
Masayuki Kuzuhara\altaffilmark{9},
Jungmi Kwon\altaffilmark{23},
Taro Matsuo\altaffilmark{24},
Shoken Miyama\altaffilmark{25},
Jun-Ichi Morino\altaffilmark{14},
Amaya Moro-Martin\altaffilmark{26},
Tetsuo Nishimura\altaffilmark{13}, 
Tae-Soo Pyo\altaffilmark{13},
Eugene Serabyn\altaffilmark{27},
Takuya Suenaga\altaffilmark{10},
Hiroshi Suto\altaffilmark{14}, 
Ryuji Suzuki\altaffilmark{14},
Yasuhiro Takahashi\altaffilmark{23,28},
Naruhisa Takato\altaffilmark{13}, 
Hiroshi Terada\altaffilmark{13},
Christian Thalmann\altaffilmark{29},
Daigo Tomono\altaffilmark{13},
Edwin L. Turner\altaffilmark{18,30},
Makoto Watanabe\altaffilmark{31},
John P. Wisniewski\altaffilmark{15},
Toru Yamada\altaffilmark{32},
Hideki Takami\altaffilmark{13}, 
Tomonori Usuda\altaffilmark{13},
\and
Motohide Tamura\altaffilmark{23,13}}

\affil{$^1$Steward Observatory, The University of Arizona, 933 N Cherry Ave, Tucson, AZ 85721, USA}
\affil{$^2$Eureka Scientific, 2452 Delmer, Suite 100, Oakland CA 96002, USA}
\affil{$^3$Department of Physics, University of Cincinnati, Cincinnati OH 45221, USA}
\affil{$^4$Guest Observer, Infrared Telescope Facility}
\affil{$^5$Also Space Science Institute, Boulder}
\affil{$^6$Leiden Observatory, Leiden University, P.O. Box 9513, 2300 RA Leiden, The Netherlands}
\affil{$^7$Institute of Astronomy and Astrophysics, Academia Sinica, P.O. Box 23-141, Taipei 106, Taiwan} 
\affil{$^8$NASA Goddard Space Flight Center, Exoplanets and Stellar Astrophysics Laboratory, Code 667, Greenbelt, MD 20771, USA}
\affil{$^9$Department of Earth and Planetary Sciences, Tokyo Institute of Technology, 2-12-1 Ookayama, Meguro, Tokyo 152-8551, Japan}
\affil{$^{10}$The Graduate University for Advanced Studies (SOKENDAI), Shonan International Village, Hayama-cho, Miura-gun, Kanagawa 240-0193, Japan}
\affil{$^{11}$Graduate School of Science, Osaka University, 1-1 Machikaneyama, Toyonaka, Osaka 560-0043}
\affil{$^{12}$The Aerospace Corporation, Los Angeles, CA 90009, USA}
\affil{$^{13}$Subaru Telescope, 650 North AÕohoku Place, Hilo, HI 96720, USA}
\affil{$^{14}$National Astronomical Observatory of Japan, 2-21-1 Osawa, Mitaka, Tokyo 181-8588, Japan}
\affil{$^{15}$H. L. Dodge Department of Physics and Astronomy, University of Oklahoma, 440 West Brooks St Norman, OK 73019, USA}
\affil{$^{16}$Laboratoire Lagrange, UMR7293, Universit\'{e} de Nice-Sophia Antipolis, CNRS, Observatoire de la C\^{o}te d'Azur, 28 avenue Valrose, 06108 Nice Cedex 2, France}
\affil{$^{17}$Max Planck Institute for Astronomy, K\"{o}nigstuhl 17, 69117 Heidelberg, Germany} 
\affil{$^{18}$Department of Astrophysical Sciences, Princeton University, Princeton, NJ, USA}
\affil{$^{19}$Department of Physics and Astronomy, College of Charleston, 66 George St., Charleston, SC 29424, USA} 
\affil{$^{20}$Universitats-Sternwarte Munchen, Ludwig-Maximilians-Universitat, Scheinerstr. 1, 81679 Munchen, Germany}
\affil{$^{21}$Institute for Astronomy, University of Hawaii, 640 North A'ohoku Place, Hilo, HI 96720, USA}
\affil{$^{22}$Department of Astronomy, Stockholm University, AlbaNova University Center, 10691 Stockholm, Sweden}
\affil{$^{23}$Department of Astronomy, University of Tokyo, 7-3-1 Hongo, Bunkyo, Tokyo 113-0033, Japan}
\affil{$^{24}$Department of Astronomy, Kyoto University, Kitashirakawa-Oiwake-cho, Sakyo-ku, Kyoto, 606-8502, Japan}
\affil{$^{25}$Hiroshima University, 1-3-2 Kagamiyama, Higashi-Hiroshima, 739-8511, Japan}
\affil{$^{26}$Departamento de Astrof\'{i}sica, CAB (INTA-CSIC), Instituto Nacional de T\'{e}cnica Aeroespacial, Torrej\'{o}nde Ardoz, 28850, Madrid, Spain}
\affil{$^{27}$Jet Propulsion Laboratory, California Institute of Technology, Pasadena, CA, 91109 USA} 
\affil{$^{28}$MEXT, 3-2-2- Kasumigaseki, Chiyoda, Tokyo 100-8959}
\affil{$^{29}$Astronomical Institute "Anton Pannekoek", University of Amsterdam,Postbus 94249, 1090 GE, Amsterdam, The Netherlands} 
\affil{$^{30}$Kavli Institute for Physics and Mathematics of the Universe,The University of Tokyo, 5-1-5, Kashiwanoha, Kashiwa, Chiba 277-8568, Japan}
\affil{$^{31}$Department of Cosmosciences, Hokkaido University, Sapporo 060-0810, Japan}
\affil{$^{32}$Astronomical Institute, Tohoku University, Aoba, Sendai 980-8578, Japan}

\begin{abstract}
We present the first resolved near infrared imagery of the transition disk Oph IRS 48 (WLY 2-48), which was recently observed with ALMA to have a strongly asymmetric sub-millimeter flux distribution. H-band polarized intensity images show a $\sim$60AU radius scattered light cavity with two pronounced arcs of emission, one from Northeast to Southeast and one smaller, fainter and more distant arc in the Northwest. K-band scattered light imagery reveals a similar morphology, but with a clear third arc along the Southwestern rim of the disk cavity.  This arc meets the Northwestern arc at nearly a right angle, revealing the presence of a spiral arm  or local surface brightness deficit in the disk, and explaining the East-West brightness asymmetry in the H-band data.  We also present 0.8-5.4$\mu$m IRTF SpeX spectra of this object, which allow us to constrain the spectral class to A0$\pm$1 and measure a low mass accretion rate of 10$^{-8.5}$M$_{\sun}$/yr, both consistent with previous estimates. We investigate a variety of reddening laws in order to fit the mutliwavelength SED of Oph IRS 48 and find a best fit consistent with a younger, higher luminosity star than previous estimates. 
\end{abstract}

\section{Introduction}

Circumstellar disks have long been hypothesized to be a breeding ground for planets, however only recently has the spatial resolution and inner working angle necessary to directly test this hypothesis been achievable. Giant planets are expected to perturb the disk as they form, changing the spatial distribution of gas and dust by opening gaps and supporting wholly or partially cleared central cavities spanning tens of AU \citep{Andrews:2011, Williams:2011}. These cavities are also detected in unresolved photometry of young stellar objects (YSOs) as ``missing'' hot thermal emission in the spectral energy distribution (SED) \citep{Strom:1989, Calvet:2005}. 

Although planet formation is a leading hypothesis for how these observed cavities are formed \citep{Zhu:2011,Dodson-Robinson:2011}, the waters are muddied by several competing, and evidently co-occurring, mechanisms for creating these gaps \citep{Alexander:2013}, which also include grain growth \citep{Dullemond:2005, Birnstiel:2012} and photoevaporation \citep{Pascucci:2009,Clarke:2001}. Each mechanism makes different predictions for the size, structure, symmetry, and depletion factor of various disk constituents (gas, large dust grains, small dust grains), thus high resolution multi-wavelength data that probe both gas and dust properties and morphologies are needed to disentangle them.  

Recent observations have revealed dust disk distortions that, unlike cleared cavities, are not conspicuous in the SEDs. These include spiral arms \citep{Muto:2012, Grady:2013, Casassus:2012, Rameau:2012}, divots \citep{Hashimoto:2011, Mayama:2012}, and asymmetries \citep{van-der-Marel:2013, Casassus:2013, Fukagawa:2013, Isella:2012, Perez:2014}.  Per current understanding of disk clearing mechanisms, these deviations from axisymmetry are most easily explicable via interactions with giant planets embedded in these disks, although internal disk dynamical processes cannot be excluded. These include gravitational instability \citep{Jang-Condell:2007}, photoelectric instability \citep{Lyra:2013}, Rossby wave instability \citep{Lyra:2013} and MRI-induced asymmetries \citep{Heinemann:2009}.

Oph IRS 48 \citep[WLY 2-48][]{Wilking:1989} is an A0$\substack{+4 \\ -1}$ \citep{Brown:2012a} star in $\rho$ Oph, with high foreground extinction \citep[A$_V$=12.9,][]{McClure:2010} at d=121 pc \citep{Loinard:2008}. The first resolved images of this disk are from \citet{Geers:2007}, who imaged it at five MIR bands from 8.6 to 18.7$\mu$m with VISIR on the VLT. While their shorter wavelength PAH images (7.5-13$\mu$m) all peak at the star, though with varying degrees of spatial extension, the 18.7$\mu$m thermal emission reveals a ring-like structure peaking at 55AU from the star. They attribute the lack of a strong 10$\mu$m silicate feature in the spectrum of this object to clearing of small $\mu$m sized dust inside the cavity seen at 18.7$\mu$m, though they note that centrally peaked and spatially extended PAH emission at other MIR bands indicates that the cavity cannot be wholly cleared of material. Indeed, subsequent modeling of PAHs in Oph IRS 48 by \citet{Maaskant:2014} find that the spectrum is dominated by highly ionized PAHs in optically thin regions inside of the disk cavity, and that the neutral PAH contribution from the outer disk is very low.

\citet{Brown:2012a} obtained high resolution (R$\sim$ 100,000) CRIRES spectra of Oph IRS 48 at the 4.7$\mu$m CO fundamental rovibrational band, and found a $\sim$30 AU ring, which they interpreted as arising from a dust wall at this radius. They modeled the excitation of the M-band CO emission and found a good fit to the optical spectrum with a 2.0$\pm$0.5 solar mass star with L=14.3 L$_\sun$. 

Subsequent ALMA 0.44mm observations of Oph IRS 48 \citep{van-der-Marel:2013} revealed that the sub-millimeter dust distribution is highly asymmetric, and is concentrated in an arc subtending  $\sim$100$^\circ$ on the south side of the star, with a factor of $>$130 contrast between that structure and the north side of the disk, where no emission is detected. $^{12}$CO 6-5 data, on the other hand, are consistent with a continuous, rotating Keplerian disk. Though a slight North/South asymmetry was noted in this data and in \citet{Brown:2012}, potential effects of foreground absorption called its reality into question. \citet{van-der-Marel:2013} interpreted the Southern 0.44mm structure as a dust trap associated with an asymmetric pressure bump,  potentially caused by a 10 Jupiter mass body located at 17$<$r$<$20 AU.  They note that smaller dust grains are less efficiently trapped under such a model \citep{Birnstiel:2013}, and are expected to be distributed more axisymmetrically about the star. Furthermore, warm H$_{2}$CO emission was detected in the southern part of the disk \citep{van-der-Marel:2014}, hinting at either a density or temperature contrast between the north and the south.

A subsequent study by \citet{Bruderer:2014} conducted detailed gas modeling on these data. These models point to an axisymmetric disk with two distinct regions inside of the observed 18.7$\mu$m cavity - one with r$<$20AU in which the gas is heavily depleted and one with 20AU$<$r$<$60AU in which the gas surface density is only marginally depleted relative to the outer disk. This complex depletion of the gas surface density inside of the disk cavity is most consistent with clearing of the disk by one or more planetary companions.  

In this paper we report the first scattered-light imagery of the disk of Oph IRS 48 and compare it with previous observations, described above, in an effort to develop a consistent multi-wavelength picture of the system. 

\section{Observations and Data Reduction}

\begin{figure*}
\centering
\subfloat[][]{
\label{fig:IRS48ims-a}
\includegraphics[scale=0.4]{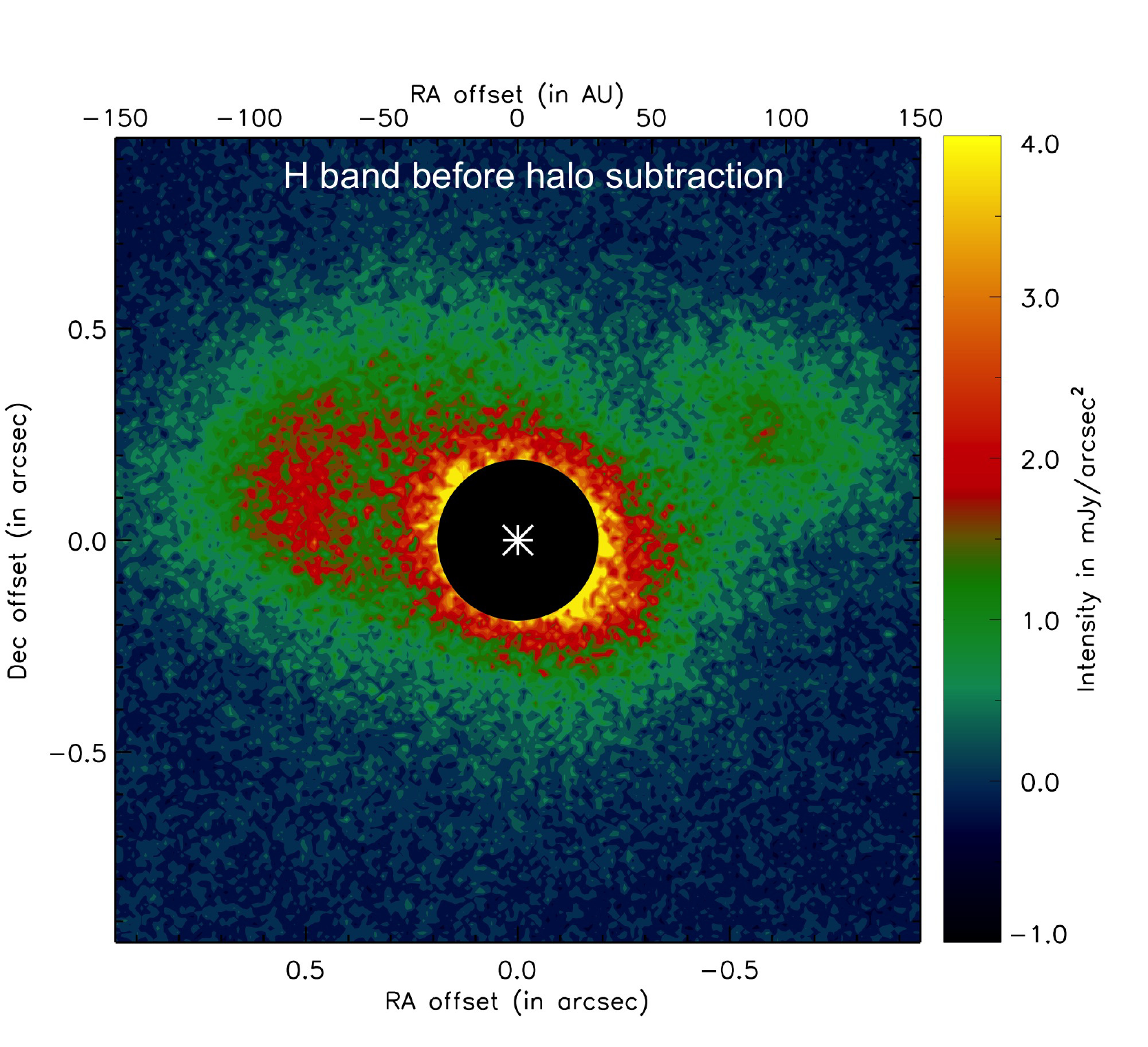}}
\subfloat[][]{
\label{fig:IRS48ims-b}
\includegraphics[scale=0.4]{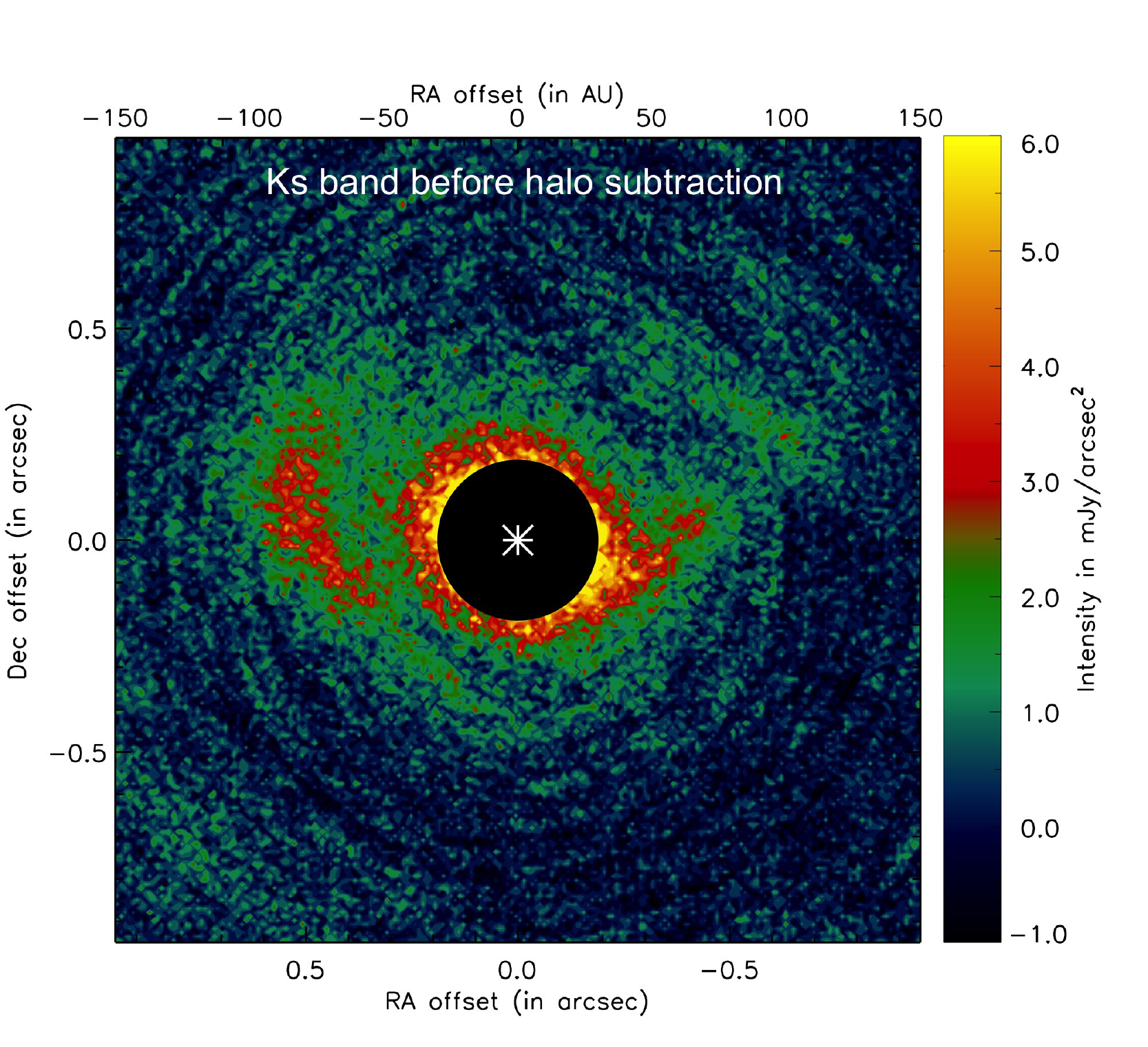}}\\
\subfloat[][]{
\label{fig:IRS48ims-c}
\includegraphics[scale=0.4]{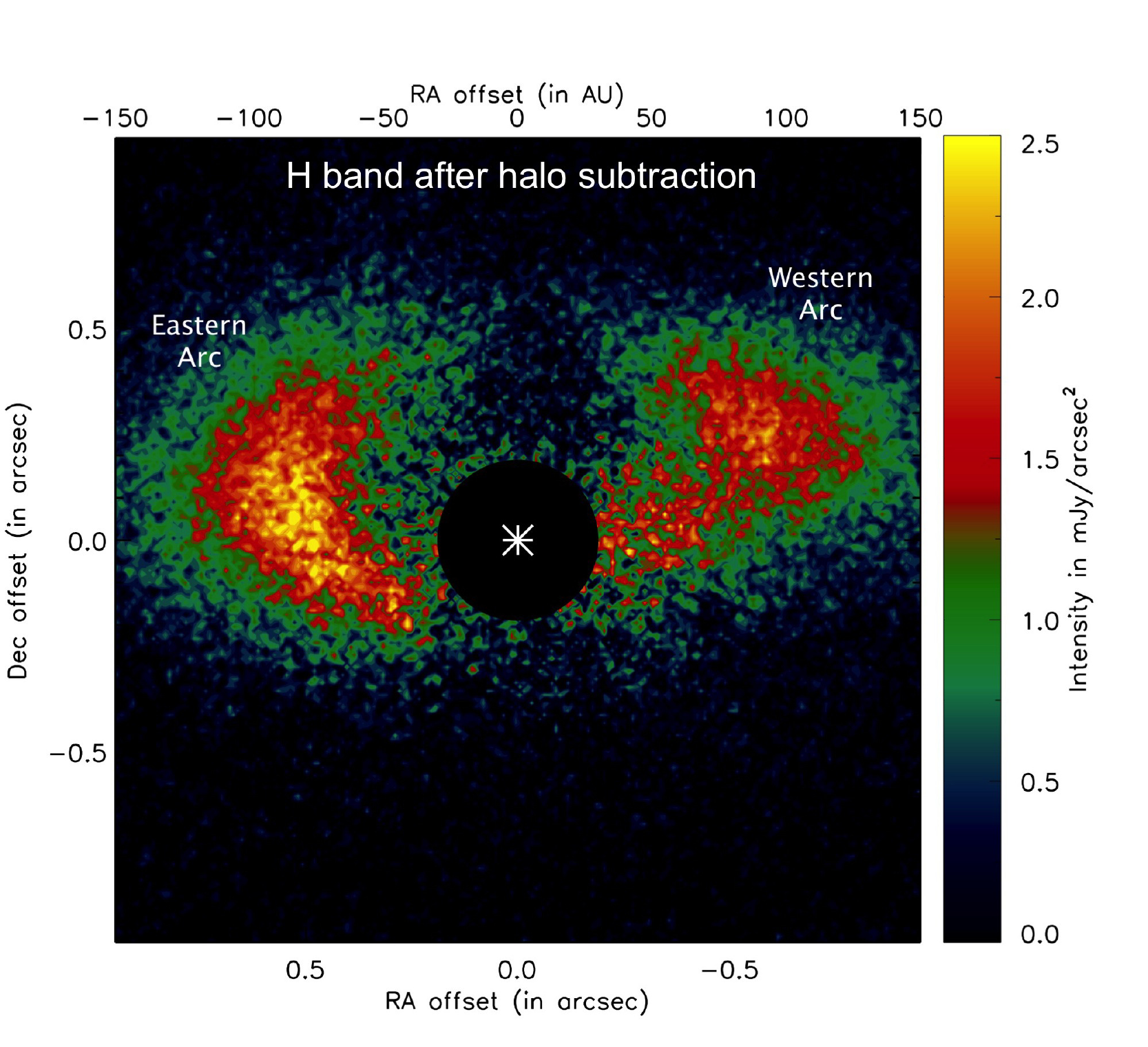}}
\subfloat[][]{
\label{fig:IRS48ims-d}
\includegraphics[scale=0.4]{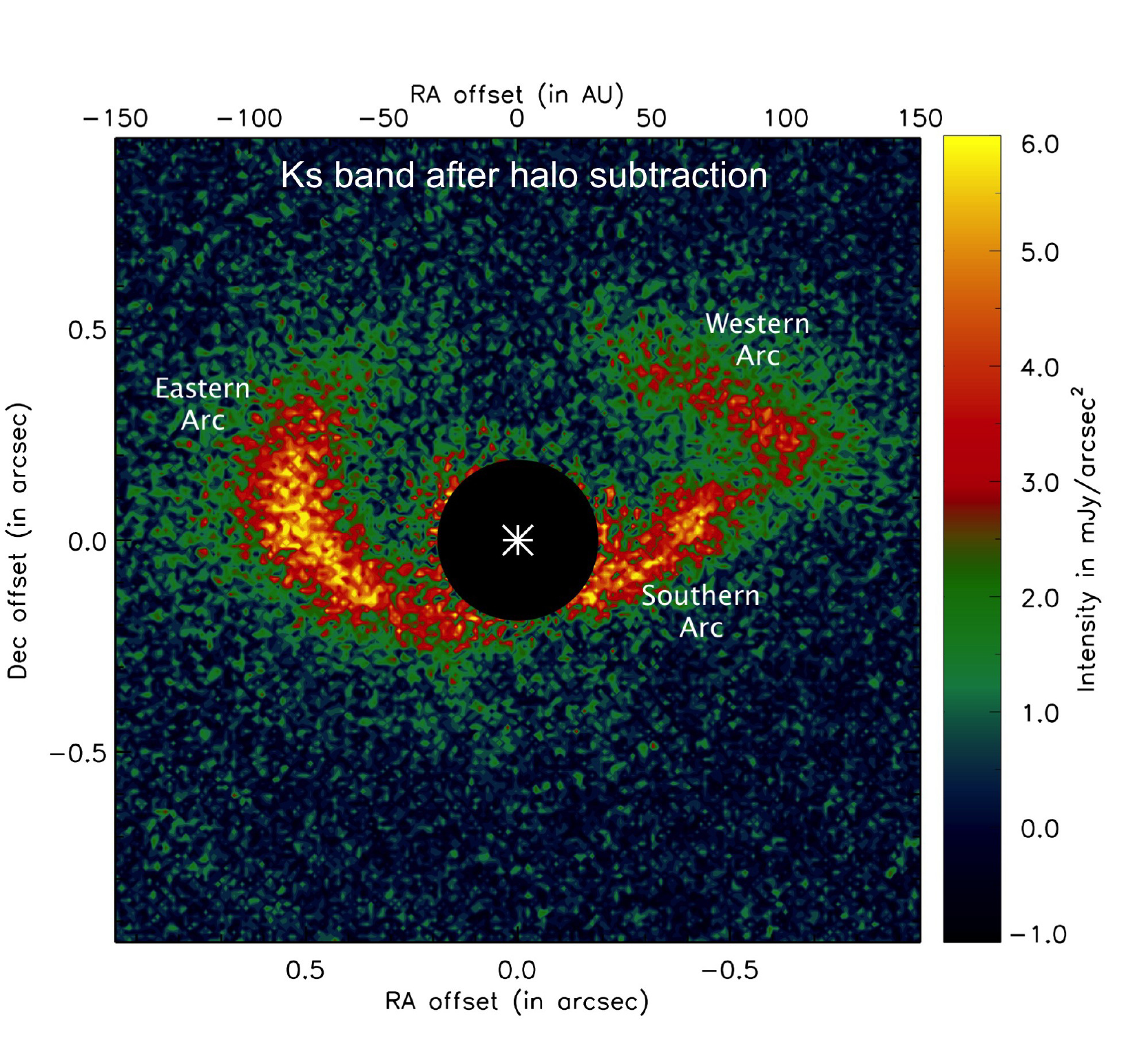}}\\
\subfloat[][]{
\label{fig:IRS48ims-e}
\includegraphics[scale=0.4]{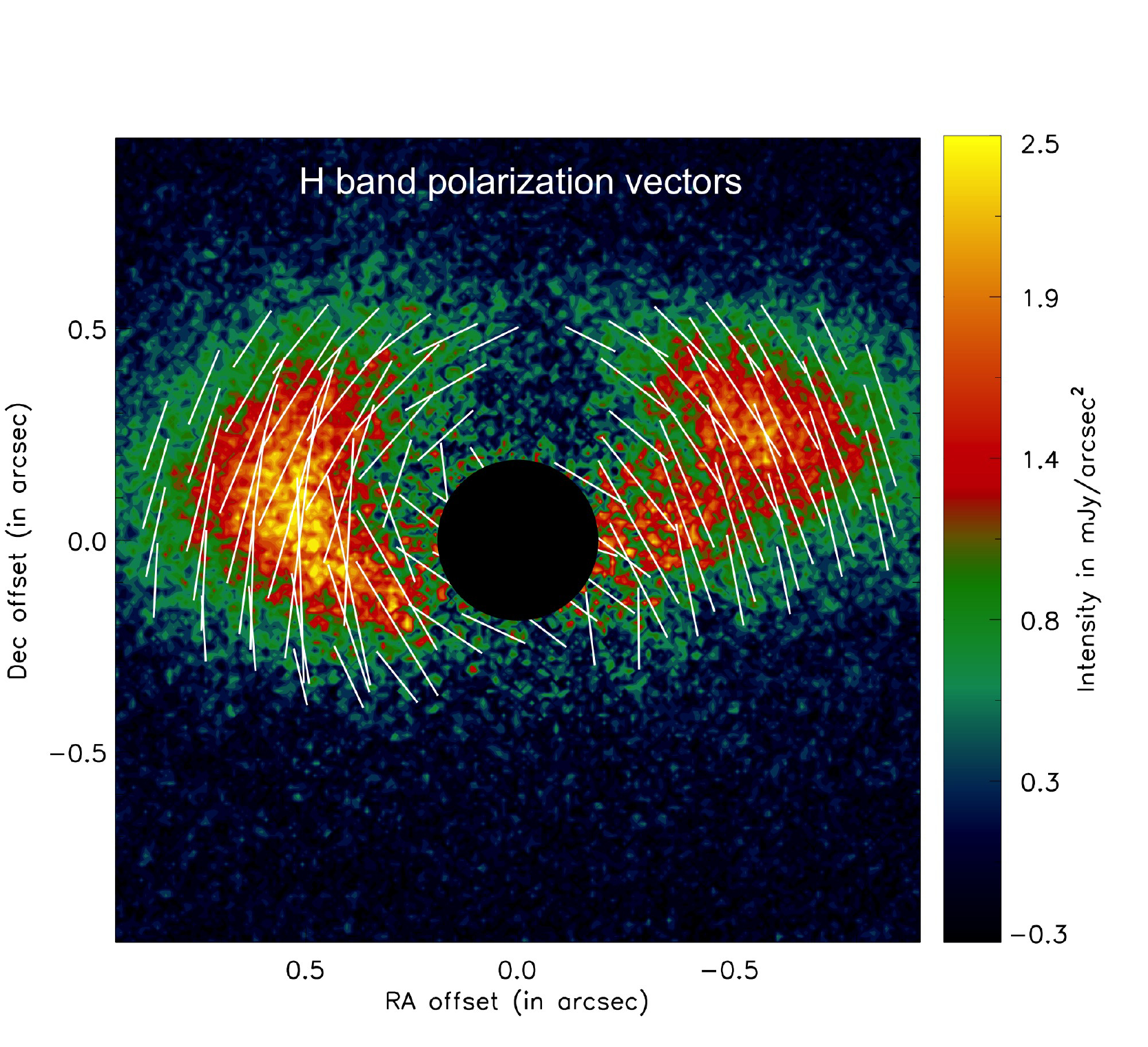}}
\subfloat[][]{
\label{fig:IRS48ims-f}
\includegraphics[scale=0.4]{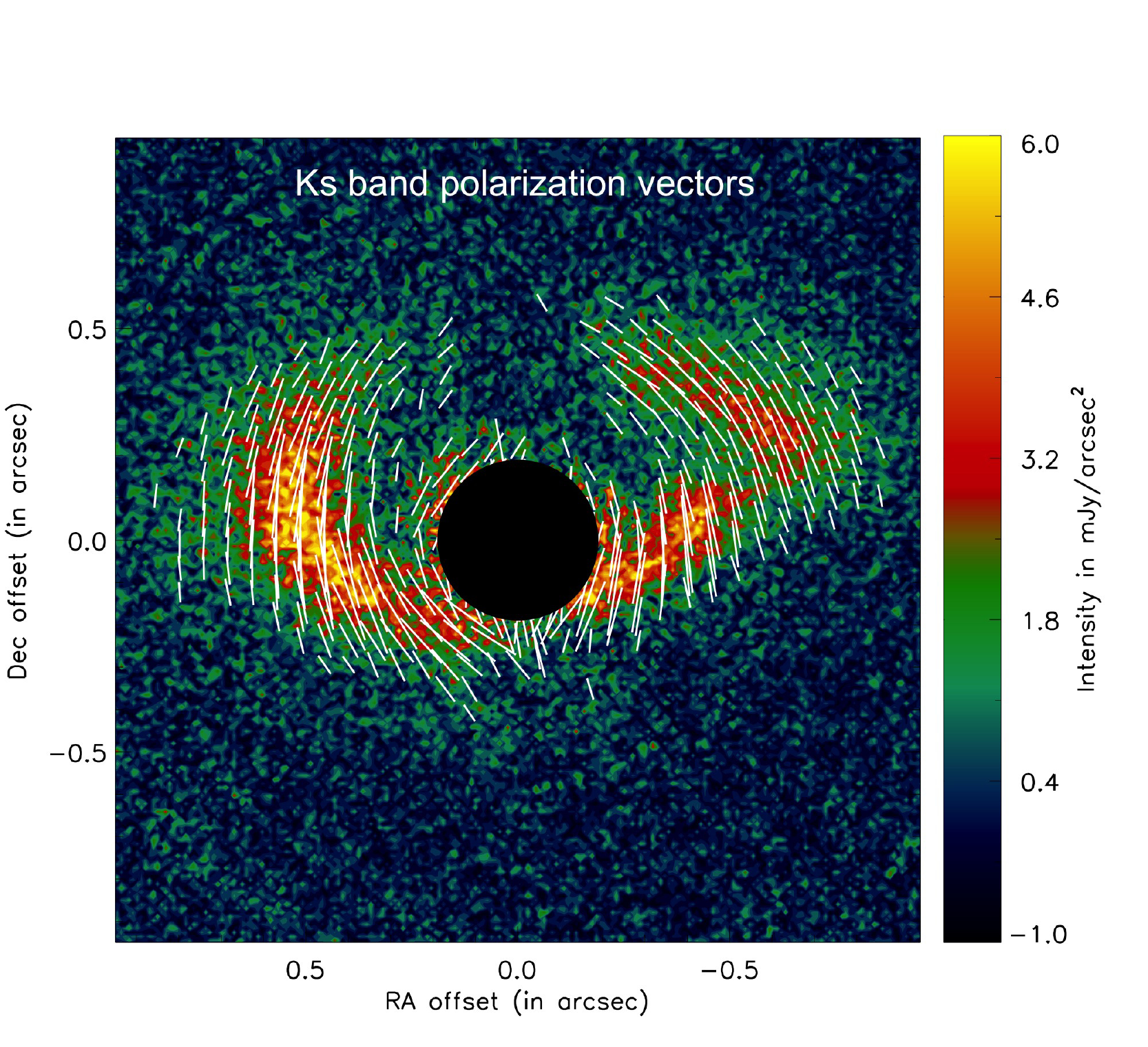}}\\
\caption[parbox=none]{Oph IRS 48 HiCIAO H-band (left) and Ks-band (right) PI images. In all cases, an r=20 (0$\farcs$19) pixel mask has been applied to cover the stellar residual. Polarization vectors are shown overplotted in bins equivalent to half of the FWHM of the star at each wavelength, and the length of each vector corresponds to the strength of the signal in that bin. Specifically, the subfigures here are:
\subref{fig:IRS48ims-a} Median H-band PI image before halo subtraction;
\subref{fig:IRS48ims-b} Median K-band PI image before halo subtraction;
\subref{fig:IRS48ims-c} Median H-band PI image after halo subtraction;
\subref{fig:IRS48ims-d} Median K-band PI image after halo subtraction;
\subref{fig:IRS48ims-e} Polarization vectors overplotted on H-band PI image after halo subtraction; and,
\subref{fig:IRS48ims-f} Polarization vectors overplotted on K-band PI image after halo subtraction.}
\label{fig:IRS48ims}
\end{figure*}

\subsection{Subaru/HiCIAO}
Polarized Differential Imaging (PDI) of Oph IRS 48 was done in H-band (1.6$\mu$m) on 2013 May 19 and in Ks band (2.2$\mu$m) on 2014 June 9 with the high-contrast imager HiCIAO \citep{Tamura:2006, Hodapp:2006} and adaptive optics system AO188 \citep{Minowa:2010} on the 8.2m Subaru telescope on Mauna Kea. These observations were conducted as part of the Strategic Exploration of Exoplanets and Disks with Subaru (SEEDS) survey, which began in October 2009. Oph IRS 48 is too faint \citep[R=16.66,][]{Erickson:2011} for use as a natural guide star (NGS), therefore a sodium laser guide star (LGS) was used to obtain high order correction, with Oph IRS 48 itself serving as the natural tip-tilt guide star. 
	
The data were taken in ``qPDI'' mode, in which a double Wollaston prism is used to split the beam into four $\sim$5$\times$5" channels, two each of o and e-polarizations. The splitting of each polarization state into two separate channels reduces saturation effects on the Hawaii 2RG detector. In order to obtain full polarization coverage and minimize artifacts, a half waveplate was rotated to four different angular positions (0$^{\circ}$, 45$^{\circ}$, 22$\fdg$5 and 67$\fdg$5) to create each PI image. This cycle was repeated 14 times at H band and 17 times at Ks band, however the Ks data included 2 waveplate cycles during a seeing spike, which have been excluded in this analysis. With 30 seconds per exposure, this translates to a total of 28 minutes of integration time at H band and 30 minutes at Ks band. 
	
Each image was bias subtracted, flat fielded and cleared of bad pixels in the standard manner for infrared data analysis using custom IDL scripts. Observations of the globular cluster M5 were then used to create a distortion solution for each channel, which was applied before the images were rotated to a common on-sky geometry. 
	 
Each individual channel was aligned using Fourier cross-correlation. These aligned images were combined to create Stokes Q and U images using standard differential polarimetry methods, namely adding together each set of two identically polarized channels and then subtracting these combinations from one another. The four channels were also added directly together to form a total intensity image. The FWHM of the star in the final, median-combined total intensity images was 0$\farcs$23 (24 pixels) at H and 0$\farcs$12 (12 pixels) at Ks. The factor of two improvement in FWHM at Ks is largely due to the better performance of the AO system at longer wavelength, as the seeing was approximately 0$\farcs$5 on both nights. 

The polarized intensity images shown in Figures \ref{fig:IRS48ims-a} (H-band) and \ref{fig:IRS48ims-b} (Ks-band) are median combinations of Polarized Intensity (PI) images computed via $PI=\sqrt{Q^{2} + U^{2}}$ for each waveplate cycle. These images suffer from a residual polarized halo as described in detail in \citet{Follette:2013A}. In the case of Oph IRS 48, which is embedded fairly deeply in the cloud (A$_{V}$=12.9), polarization due to intervening dust may also contribute significantly to this effect. Figures \ref{fig:IRS48ims-c} and \ref{fig:IRS48ims-d} show the same images after removal of the polarized halo. The halo was computed by calculating the polarization P and polarization angle $\theta_{p}$ in the region 0$<$r$<$40 pixels (0$\farcs$38) and -180$^{\circ}\negmedspace<$PA$<$-150$^{\circ}$ at H-band and 0$<$r$<$15 pixels (0$\farcs$14) and 160$^{\circ}\negmedspace<$PA$<$200$^{\circ}$ for Ks band. These regions were selected iteratively from among the regions near the star with little or no disk emission. The quality of halo subtraction was judged in each case by the strength of the stellar residual remaining at the center of the image. The halo values that minimized these stellar residuals were P=0.6\% and PA=41$^{\circ}$ at H band and P=0.5\% and PA=36$^{\circ}$ at Ks band. The halo was removed most effectively in the H-band data, while the K-band data, though significantly improved through this process, still house a fairly strong stellar residual at the center of the cavity that may be reflective of real inner disk emission.  These residuals have been masked in all images.

Polarization angles were calculated for each halo-subtracted image according to the formula  $\theta_{p}= 0.5 $arctan$ (U/Q)$ and median combined. These are shown overplotted on the PI image in 12/6 pixel (0$\farcs$11/0$\farcs$06) bins for H/Ks (half of the stellar FWHM in each case) after halo subtraction in Figures \ref{fig:IRS48ims-e} and \ref{fig:IRS48ims-f}. The length of each vector has been scaled according to the magnitude of the polarized intensity in that bin, and vectors in bins at or below the background level have been clipped. Halo subtraction results in strong centrosymmetry, as expected for polarized disk emission. 

A PSF star, HD148835, was also observed at H band on 2013 May 19 just after Oph IRS 48 observations were completed. Five waveplate cycles of 15 second images were completed for this brighter star for a total of 5 minutes of integration time. The same reduction procedure was followed for this data set in order to verify that nothing mimicking disk structure was induced by the halo subtraction procedure. In the case of the PSF star, halo subtraction both reduces the strength of the stellar residual and randomizes the direction of the polarization vectors, as expected for a star without any circumstellar material. 

We also attempted classical PSF subtraction using the total intensity images of Oph IRS 48(V=17.7, R=16.7) and HD148835 (V=10.07, R=9.26), but found the PSFs to be poorly matched and were not able to isolate disk emission in this way. The large difference in brightness between the two stars should result in significantly different AO performance and therefore PSF structures, so this poor match is to be expected.  

\begin{figure}[b]
\centering
\subfloat[][]{
\label{fig:SpeX-a}
\includegraphics[scale=0.48]{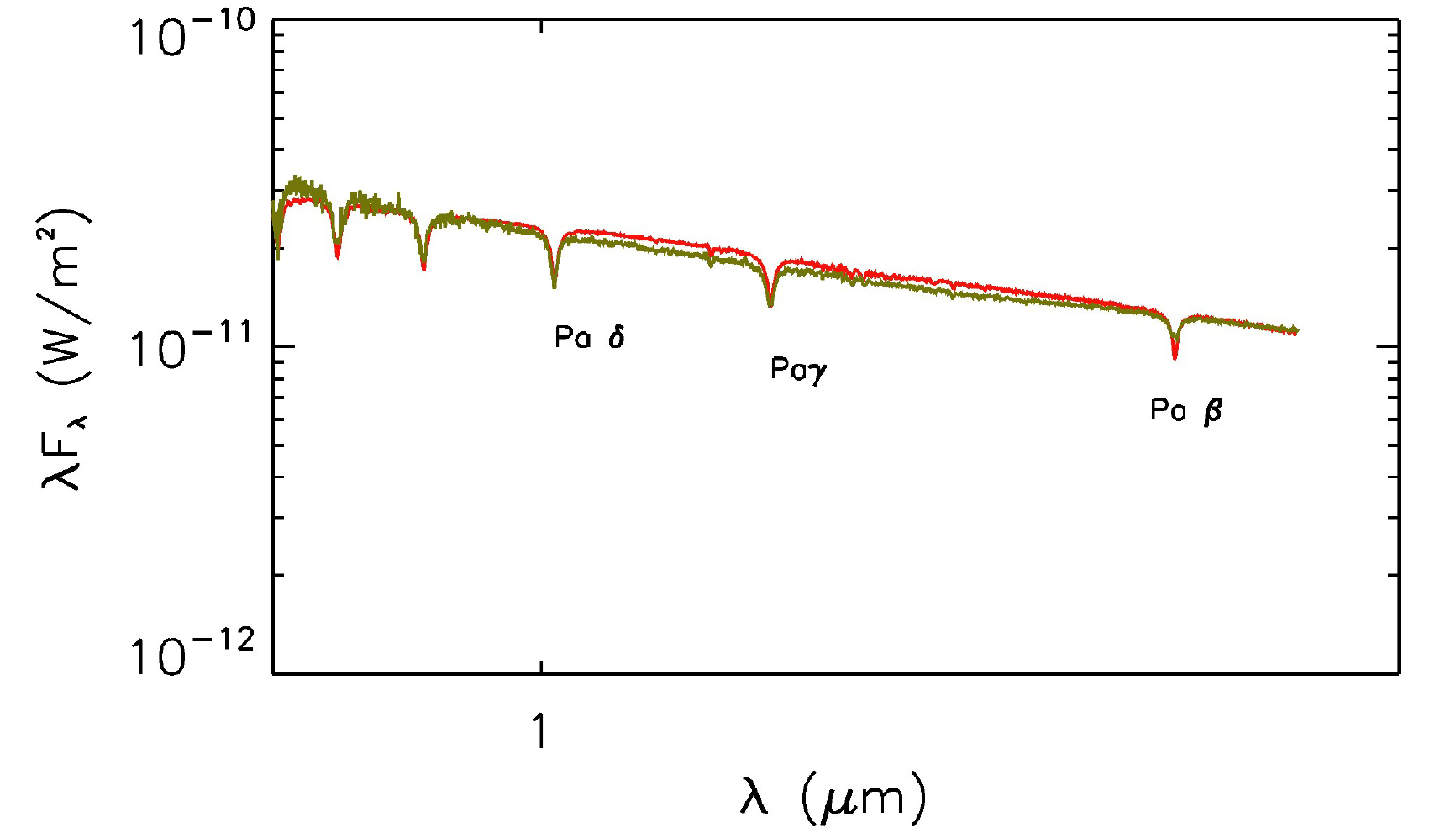}}\\
\subfloat[][]{
\label{fig:SpeX-b}
\includegraphics[scale=0.47]{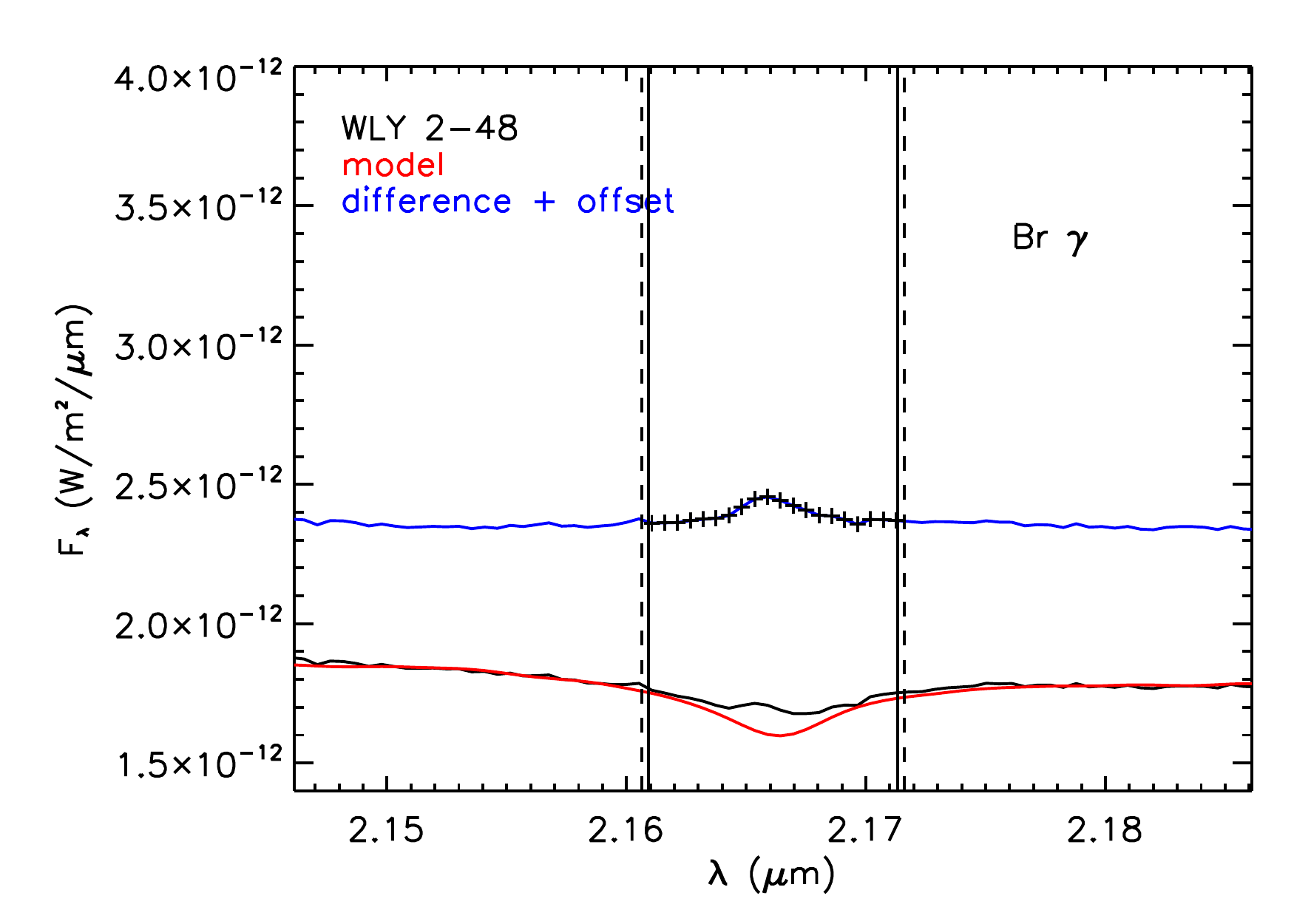}}\\
\subfloat[][]{
\label{fig:SpeX-c}
\includegraphics[scale=0.5]{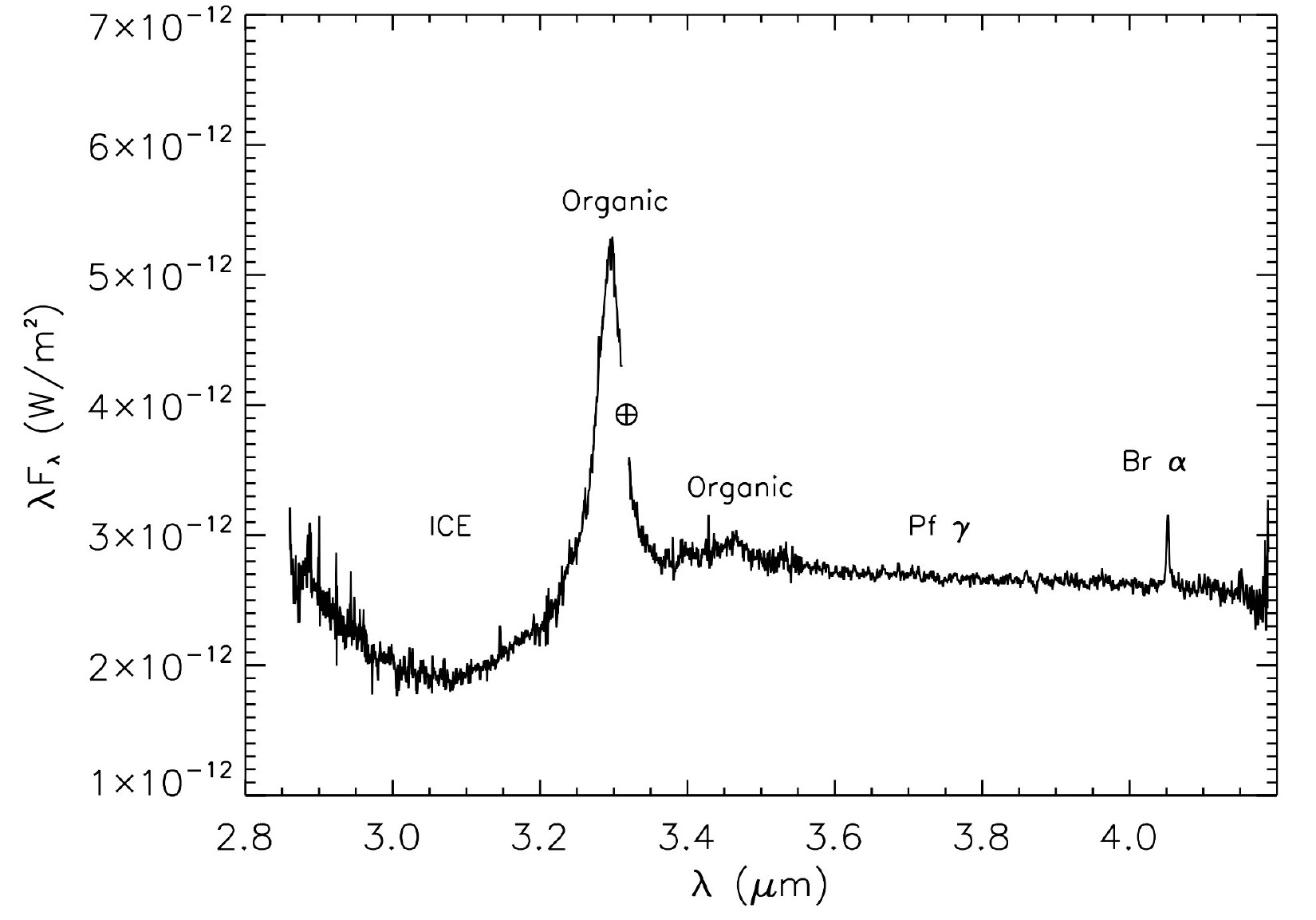}}
\caption[]{
\subref{fig:SpeX-a} 0.8-2.4$\mu$m SpeX spectrum, with the spectrum of another A0 star, SAO206463 overplotted.; \subref{fig:SpeX-b} Pa$\beta$ line fit used together with a similar fit to the Br$\gamma$ line to derive a weak accretion rate of $\dot{M}$=10$^{-8.5}$ M$_{\sun}$/yr for Oph IRS 48.;
\subref{fig:SpeX-c} 2.3-4.3$\mu$m SpeX spectrum, with pertinent lines labeled.}
\label{fig:SpeX}
\end{figure}

\begin{figure*}[!htbp]
\centering
\subfloat[][]{
\label{fig:ellipse-a}
\includegraphics[scale=0.4]{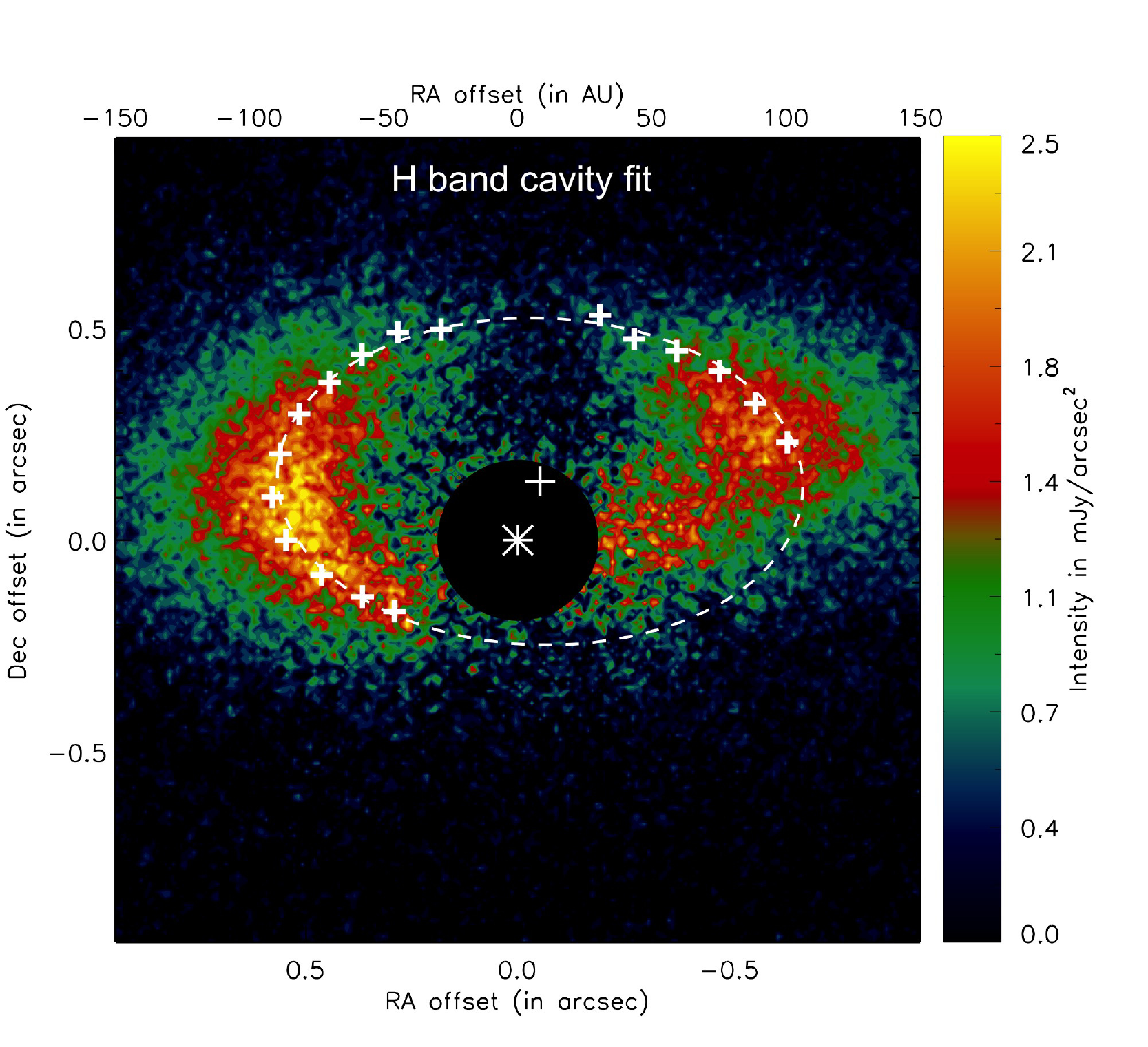}}
\hspace{8pt}
\subfloat[][]{
\label{fig:ellipse-b}
\includegraphics[scale=0.4]{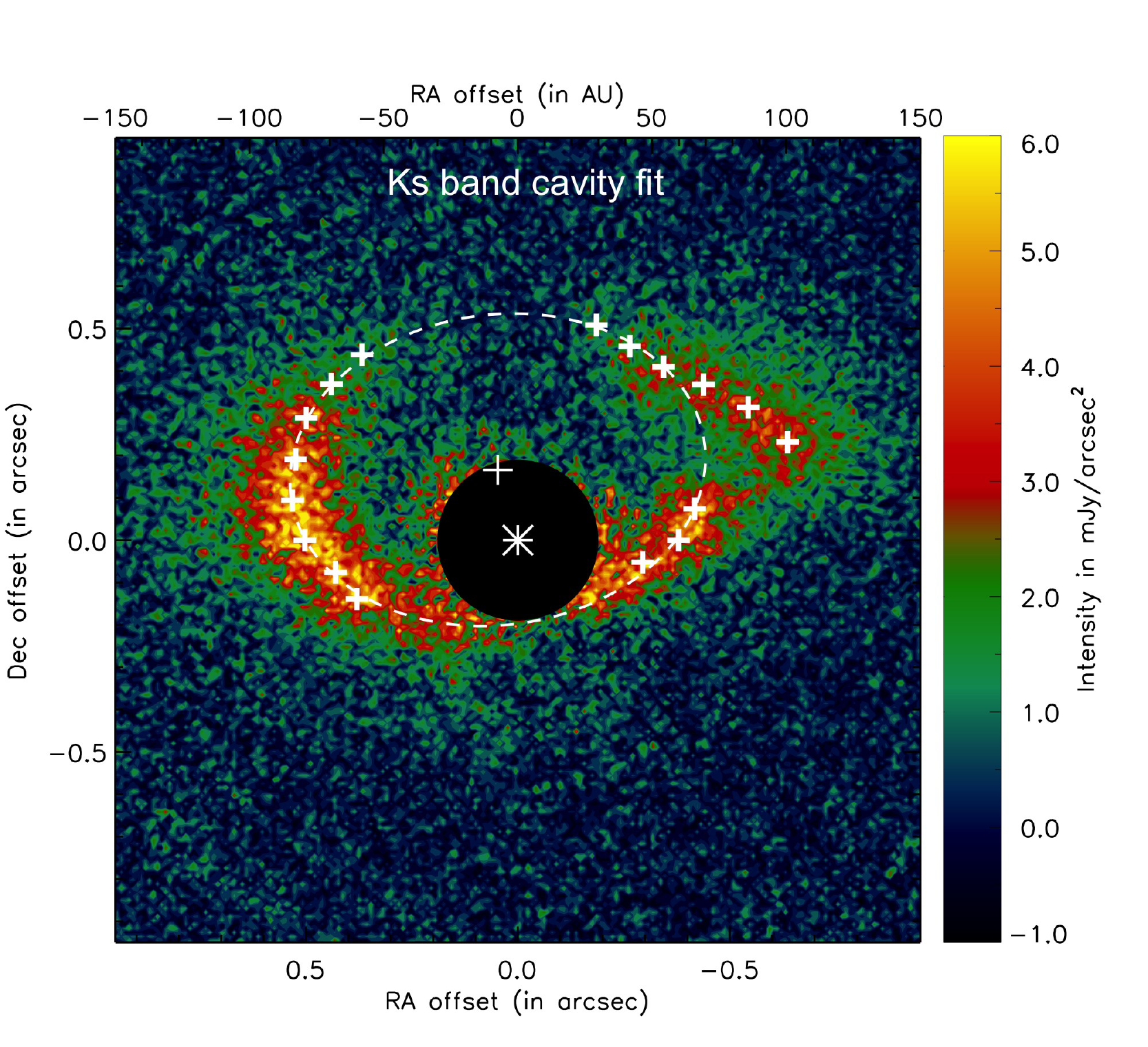}}\\
\subfloat[][]{
\label{fig:ellipse-c}
\includegraphics[scale=0.4]{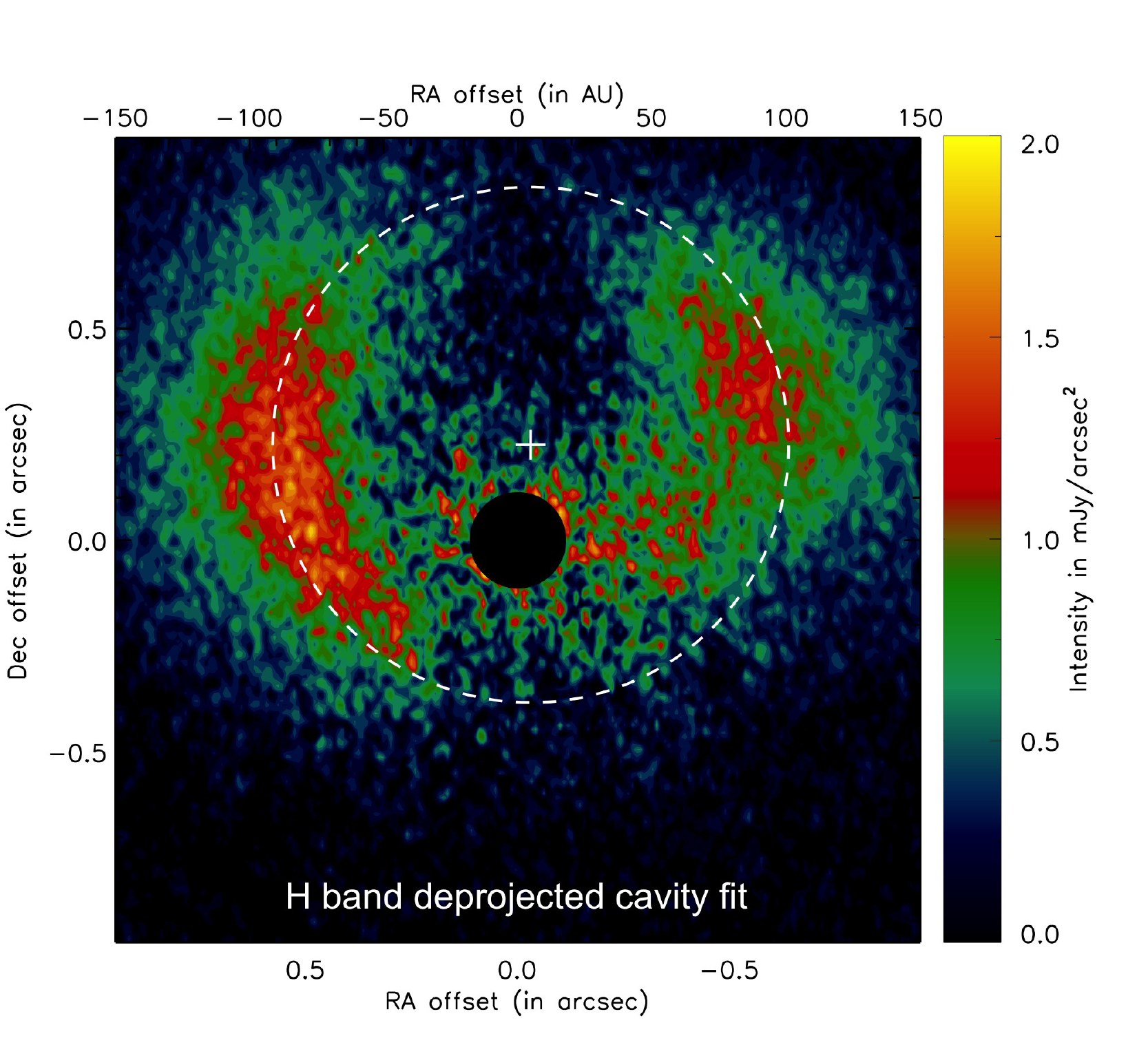}}
\hspace{8pt}
\subfloat[][]{
\label{fig:ellipse-d}
\includegraphics[scale=0.4]{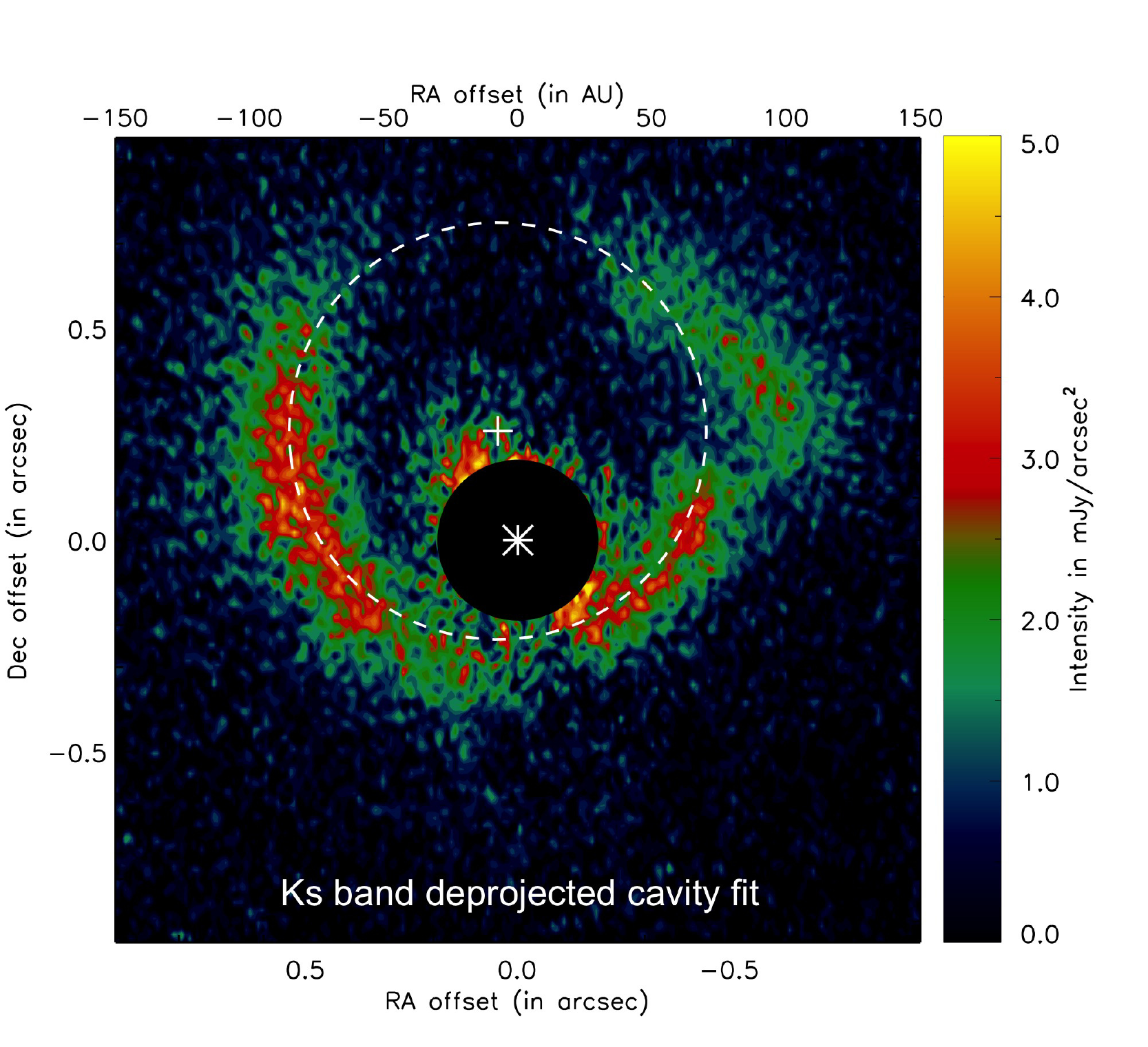}}\\
\caption{Top panels: Elliptical fits to the Oph IRS 48 scattered light cavity at H (left) and Ks (right) bands. The locations of the peaks for each 10$^{\circ}$ azimuthally binned radial profile are shown with white crosses. The best fit ellipse to these points is shown overplotted as a dashed white line in each case. The center of that ellipse is shown with a large white cross. As discussed in the text, the Ks band cavity fit is significantly more robust, as it is (i) higher resolution data, (ii) brighter and (iii) the morphology is clear enough at this wavelength that profiles dominated by the asymmetry can be excluded. Bottom panels: The same images, but deprojected along PA=97.4$^{\circ}$ to account for an inclination of 50$^{\circ}$ and with the deprojected circular cavity fit shown. The large northern offset between the stellar centroid and the cavity center is discussed in detail in Section \ref{offset}}
\label{fig:ellipse}
\end{figure*}
\subsection{IRTF/SpeX}
We obtained NIR spectra of Oph IRS 48 on 2013 May 16 using the SpeX \citep{Rayner:2003}) spectrograph on NASA's Infrared Telescope Facility (IRTF). The spectra were recorded using the echelle grating in both short-wavelength mode (SXD, 0.8-2.4 \micron{}, Figure \ref{fig:SpeX-a}) and long-wavelength mode (LXD, 2.3-5.4 \micron{}, Figure \ref{fig:SpeX-b}) using a 0$\farcs$8 slit. The spectra were corrected for telluric extinction and flux calibrated against the A0V calibration star HD 145127 using the Spextool data reduction package \citep{Vacca:2003,Cushing:2004}. In addition to the 0$\farcs$8-slit spectra, we also recorded data with the SpeX prism disperser and a wide 3.0" slit, which allows us to retrieve the absolute flux levels when the sky transparency is good and the seeing is 1" or better (it was 0$\farcs$6 arcsec at the time the data were acquired).

\section{Results}

In this section, we present the results of our analysis of the new HiCIAO and SpeX data. In Section 3.1, we discuss the geometry of the disk in scattered light.  In Section 3.2, we present fits to the Oph IRS 48 cavity. In section 3.3, we discuss the azimuthal symmetry, or lack thereof, in the data based on radial profiles through the deprojected disk image. In section 3.4, we present a revised age and spectral type derived from our analysis of the multiwavelength SED, and in section 3.5, we present accretion estimates derived from our SpeX spectra.  

\subsection{Observed Disk Geometry}

The disk as seen in PI extends approximately 1$\farcs$2$\pm$0$\farcs$2 ($\sim$145$\pm$25AU) from the star at H-band and 1$\farcs$0$\pm$0$\farcs$1 ($\sim$120$\pm$12AU) at Ks-band. This is similar to the extent of 12CO gas emission traced by ALMA \citep[$\sim$160AU]{Bruderer:2014}, as expected for small grains, which should be tightly coupled to the gas in the disk. Emission from the cavity rim begins $\sim$0$\farcs$3 ($\sim$20AU) from the center of the cavity and peaks at $\sim$0$\farcs$5$\pm$0$\farcs$1 (55$\pm$15AU) at most azimuths for both wavelengths. Disk parameters derived from fits to this cavity are described in detail in Section \ref{cavmorph}.

The scattered light emission from this disk has marked asymmetries both in the North/South and East/West directions that cannot be explained purely by inclination effects, as discussed in Section \ref{asymm}. In particular, the H-band emission is concentrated in two arcs, which we will call the ``Eastern" and ``Western" arcs from here forward. The Eastern arc extends for $\sim$100$^{\circ}$ from the Northeast to the Southeast of the star (20$<$PA$<$120). The Western arc is much shorter, and lies entirely northward of the stellar position, extending just 40$^{\circ}$ along PAs -70$<$PA$<$-40. 

With one notable exception, the morphology of the K-band PDI data are similar to the H-band data. The location and extent of emission along the Eastern and Western arcs is similar between the two image sets, however the Ks-band data reveal a third arc in the South that is not readily visible at H-band along -150$^{\circ}<$PA$<$-70$^{\circ}$. The morphology of this arc relative to the others, and its implications for disk structure and composition are discussed in detail in section \ref{spiral}. 

\begin{figure}
\includegraphics[scale=.3]{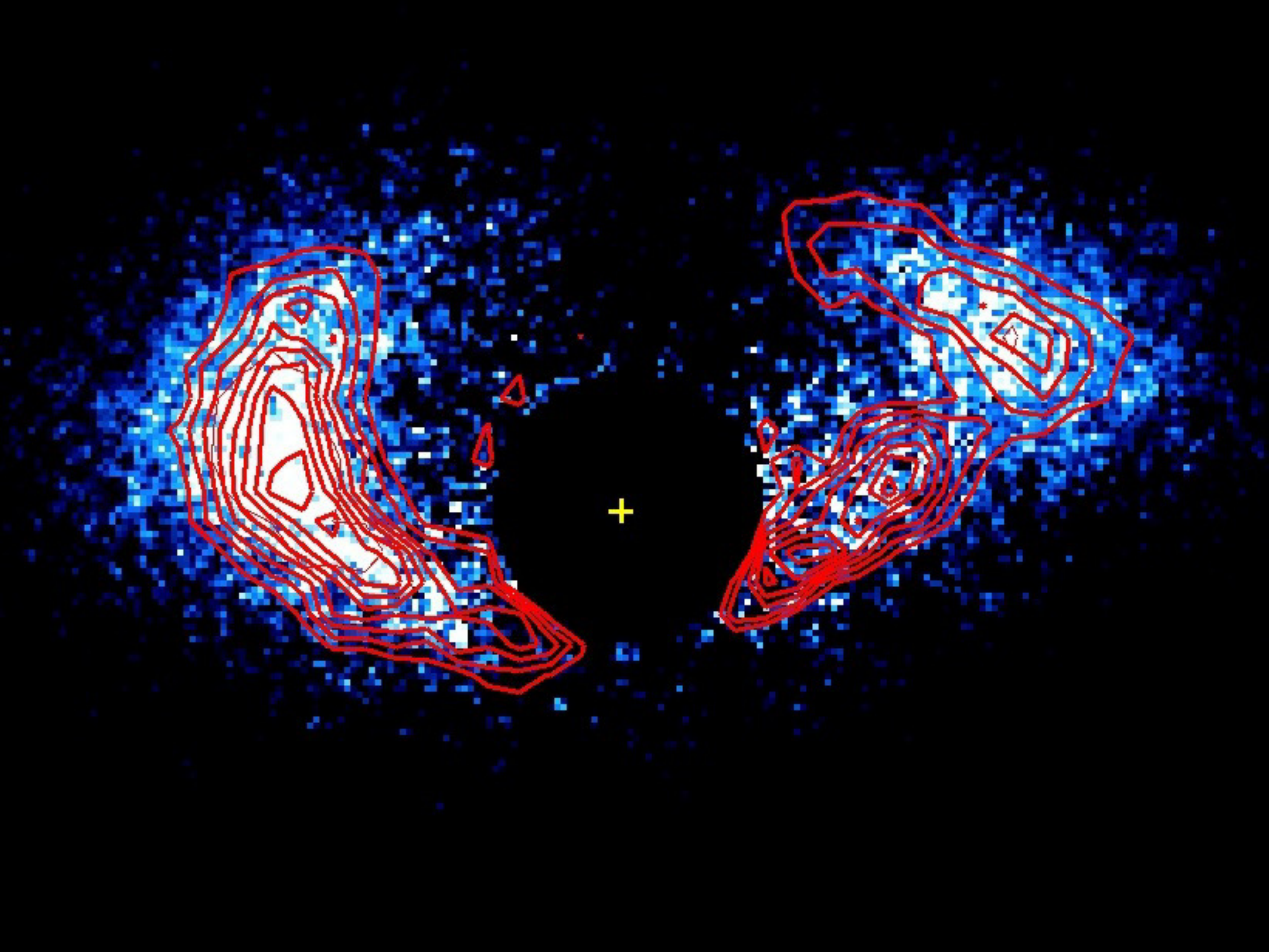}
\caption{Ks band polarized intensity contours overplotted on the H band polarized intensity image of OPH IRS 48. The improvement in image quality in the K band image clarifies that the excess emission seen to the West of the star at H band is probably the Southern disk rim, clearly visible in the K band image.}
\label{fig:HKoverlay}
\end{figure}

\subsection{Cavity Morphology}
\label{cavmorph}
To asses the morphology of the Oph IRS 48 cavity in a quantitative sense, we created radial profiles binned  to half of the FWHM of the stellar PSF for each image set (12 pixels/0$\farcs$11 at H, 6 pixels/0$\farcs$06 at Ks) for each  10$^{\circ}$ azimuthal slice in the disk. We fit the peak of each radial profile with a Gaussian and inspected the quality of fit by eye. Only those profiles well fit by a Gaussian (namely those with a clear cavity rim feature present) were included. The peak location for each profile was weighted by its intensity and fit with an ellipse to produce a least-squares elliptical cavity fit. These fits are shown overplotted on the H and Ks band data in Figures \ref{fig:ellipse-a} and \ref{fig:ellipse-b} respectively.

In the Ks-band PI data, the Southern and Western arcs appear to intersect one another at nearly a right angle, and the radial distance from the star to the rim also jumps rapidly at this intersection. We interpret this as non-axisymmetric structure in the disk - either a spiral arm that extends from the Western rim out to the Southwest or a local surface brightness deficit at the cavity wall. Both scenarios are investigated in detail in Section \ref{spiral}. As revealed in the overlay in Figure \ref{fig:HKoverlay}, the majority of the Western H-band arc lies along this spiral arm, which tends to degrade the quality of the cavity fit at this wavelength, however we report its parameters here for completeness. 

The best-fit ellipse to the H-band data, shown in Figure \ref{fig:ellipse-a}, has an eccentricity of 0.78 (a=0$\farcs$62, b=0$\farcs$39), a PA of 92$^{\circ}$, and is offset relative to the stellar centroid by +6 pixels (0$\farcs$06) in RA and +15 pixels (0$\farcs$14) in Declination. 

The higher resolution of the Ks band data allows us to exclude PAs of -50$^{\circ}$, -60$^{\circ}$ and -70$^{\circ}$ from the elliptical cavity fit, since they lie along the asymmetry. This leads to an elliptical cavity that is well fit to all of the other radial profile peaks. We thus consider the fit to the Ks-band cavity to be much more robust than the H-band fit. 

The best fit ellipse at Ks-band, shown in Figure \ref{fig:ellipse-b} has an eccentricity of 0.66 (a=0$\farcs$49, b=0$\farcs$37, consistent with an inclination of 42$\pm$10$^{\circ}$), a PA of 97$\pm$2$^{\circ}$, and a center that is offset from the stellar residual by -5 pixels (0$\farcs$05) in RA and +18 pixels (0$\farcs$17) in Declination. These values are in keeping with the published values for the inclination (50$^{\circ}$) and PA (100.3$^{\circ}$) of the disk. The inclination, in particular, is not well constrained by our data, and so we adopt the literature value of 50$^{\circ}$ in deprojecting our images. The offset of the cavity center from the stellar residual is discussed in detail in Section \ref{offset}. 

The K band fit corresponds to a deprojected circular cavity size of 59$\pm$10AU at 121pc, where error bars were derived from the average width of the Gaussian fit to each radial profile weighted by the peak. This is also consistent with the radius of the cavity as seen at other wavelengths. The circular cavity fits are shown overplotted on the deprojected disk images in Figures \ref{fig:ellipse-c} and \ref{fig:ellipse-d}

The degree of depletion of small grains inside the cavity is not well constrained by our data and is severely limited by stellar residuals. The smallest ratio between the trough (inside the cavity) and peak (at cavity rim) of a radial profile in our data is 0.40 at H-band and 0.33 at Ks-band, thus the NIR-scattering grains inside of the cavity are depleted by, at minimum, a factor of 2.5-3 relative to the outer disk.  

\begin{figure*}[t]
\centering
\subfloat[][]{
\label{fig:rps-a}
\includegraphics[scale=0.45]{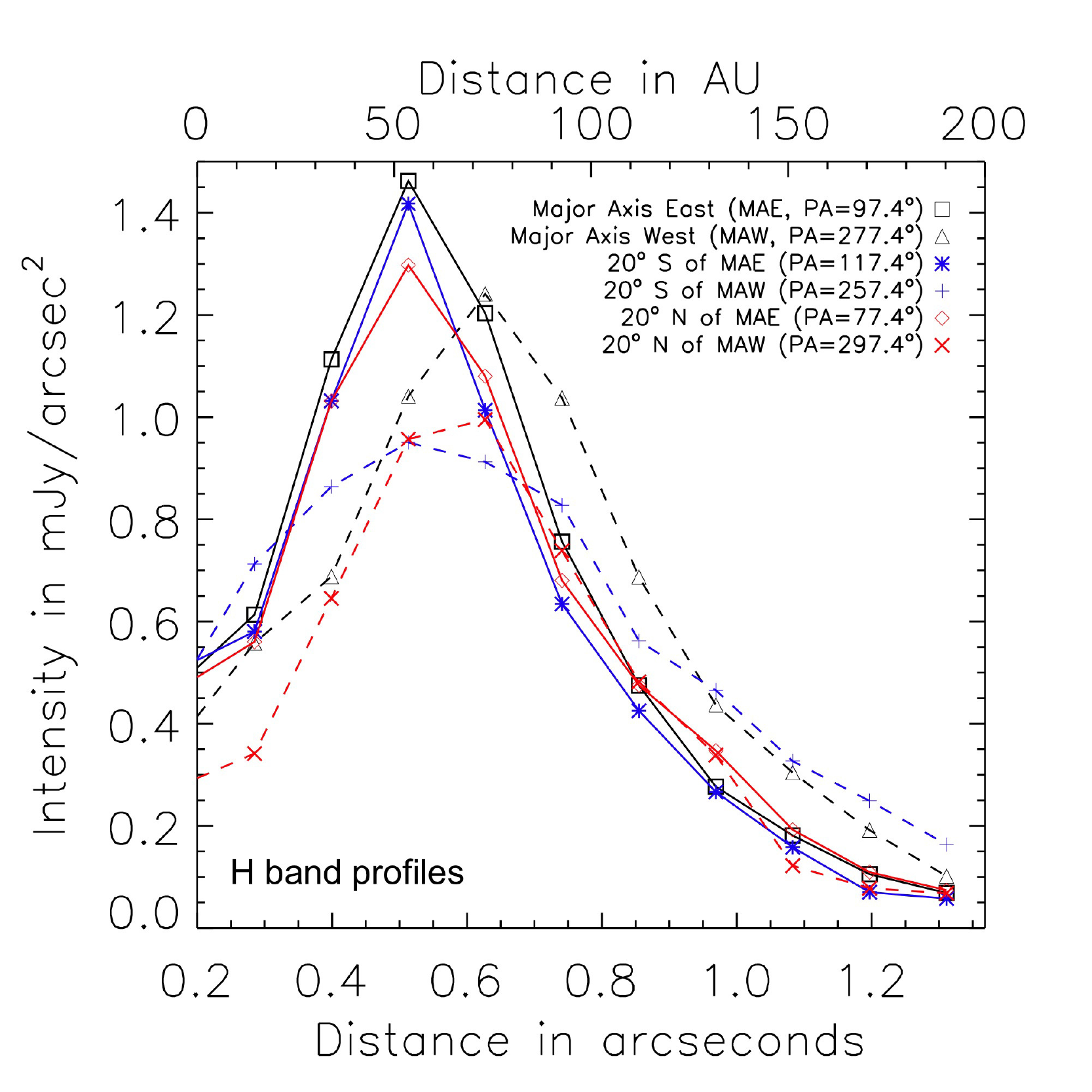}}
\subfloat[][]{
\label{fig:rps-b}
\includegraphics[scale=0.45]{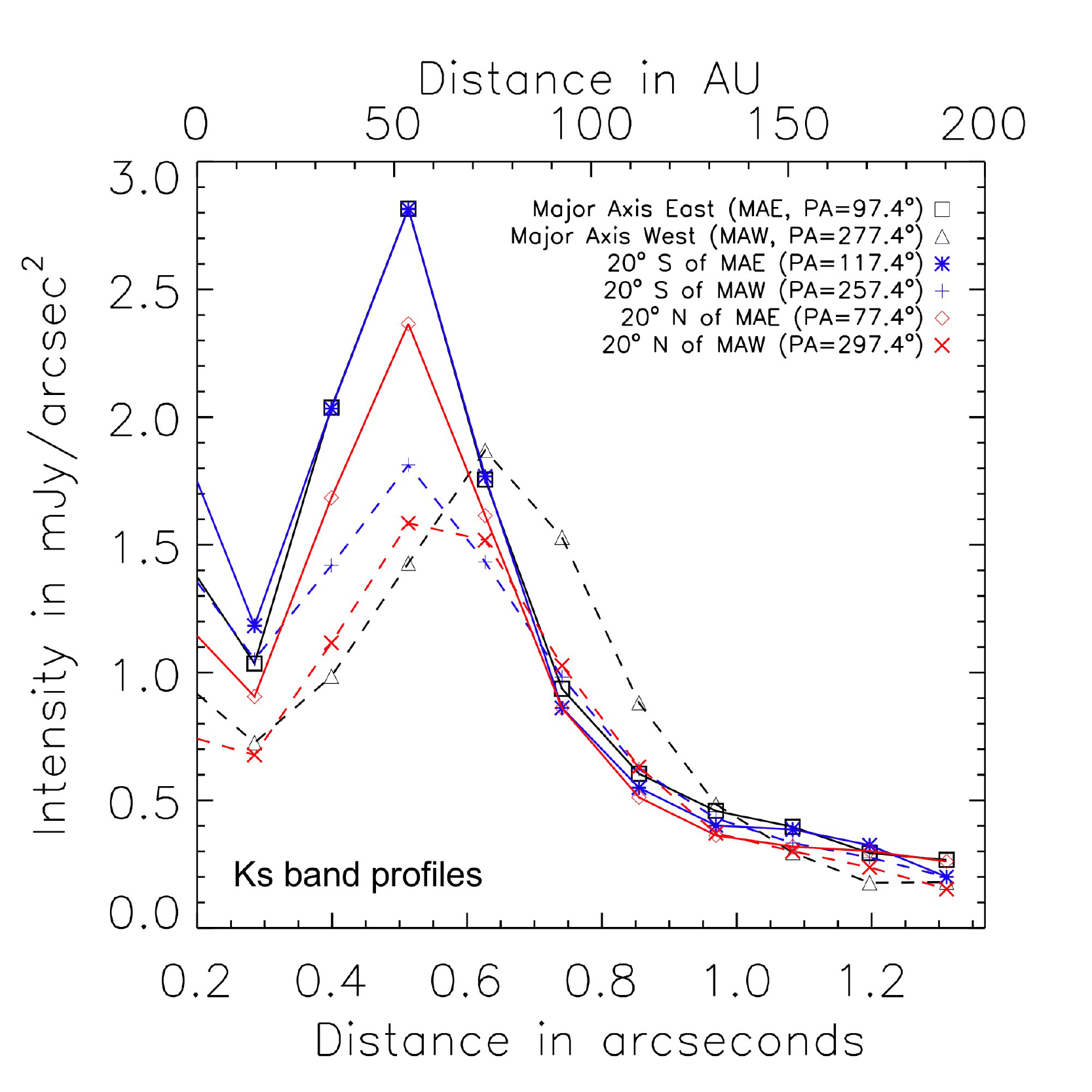}}
\label{fig:rps}
\caption[parbox=none]{Radial profiles taken through the H (left) and Ks (right) images. Azimuthal asymmetry is especially apparent along the Western major axis (dashed black lines), where profiles at both wavelengths peak $\sim$0$\farcs$15 farther from the cavity center, though the asymmetry is also hinted at in the Southwestern profiles (red dashed lines).  The blue scattering properties of the asymmetry are readily apparent - the H-band Western major axis profile peaks at $\sim$85$\%$ of the Eastern major axis value, while at Ks-band this value is $\sim$25$\%$ lower ($\sim$60$\%$). Profiles were taken with respect to the apparent cavity center rather than the location of the stellar residual, and this choice is described in detail in Appendix \ref{radprofs}. The disk images were deprojected to account for an inclination of 50$^{\circ}$ along PA=97.4$^{\circ}$, and the Ks band image was convolved with a 5.65 pixel (0$\farcs$05) Gaussian in order to simulate similar Strehl-ratio imagery. In all cases, solid lines represent profiles through the Eastern half of the disk and dashed lines represent profiles through the Western half of the disk. Black profiles were taken along the disk major axis and blue and red profiles were taken 20$^{\circ}$ North and South of the major axis, respectively.}
\end{figure*}

\subsection{Radial Profiles and Azimuthal Structure}
\label{rps}

The disk images were deprojected to remove the effect of the i=50$^{\circ}$ inclination, and allow for a direct comparison of radial profiles through the disk in any given direction. These are shown for H and Ks band in Figure \ref{fig:rps}, where in each case the disk has been deprojected along PA=97.4$^{\circ}$ for an inclination of 50$^{\circ}$. 

The profiles were taken with respect to the cavity center and not the stellar residual, and the Ks-band data were convolved with a 5.65 pixel (0$\farcs$05) Gaussian in order to simulate similar AO system performance and allow for a direct comparison of the two wavelengths. These choices are described in detail in Appendix \ref{radprofs}. 

The asymmetry in the disk is immediately apparent in Figure \ref{rps}, as the profile along the Western major axis (dashed black line) peaks $\sim$0$\farcs$15 more distant from the cavity center than the other profiles at both wavelengths. The profiles 20$^{\circ}$ South of the Western major axis are also somewhat deviant, peaking between those of the rest of the disk (at $\sim$0$\farcs$5) and the Western major axis. 

The Ks-band profiles are a factor of 2 greater in absolute intensity than the H-band profiles. This is to be expected given the region of high extinction in which Oph IRS 48 resides. Indeed the 2MASS K-band magnitude (7.582) of the star is 3 times brighter than the H-band magnitude (8.815). The profiles are generally similar, however, in relative brightness between the H and Ks band data, except in the region of the asymmetry. Deviations in this region are discussed in detail in Section \ref{spiral}.

\begin{figure*}
\includegraphics[scale=0.75]{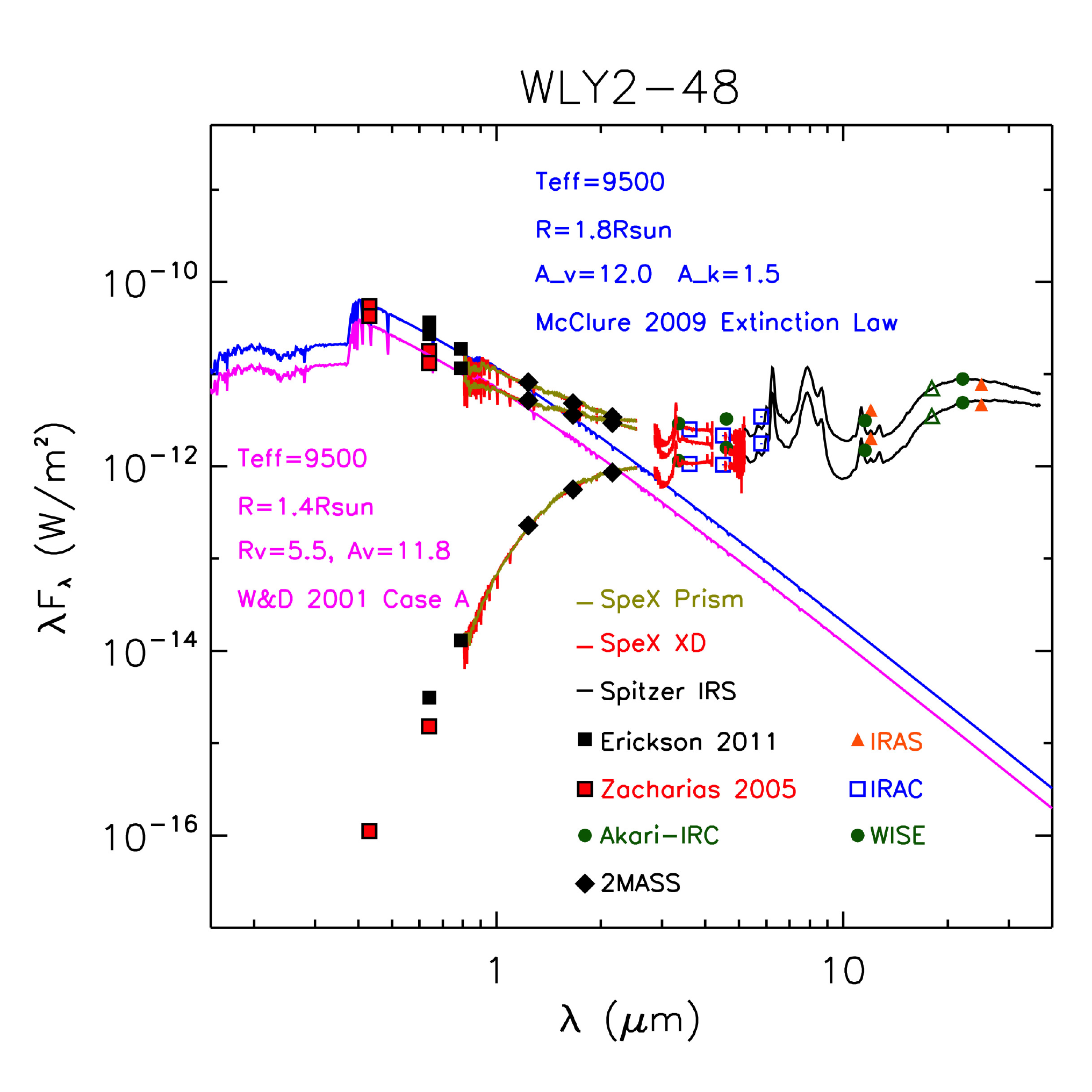}
\caption{The spectral energy distribution (SED) of Oph IRS 48. The lower set of data are observed values, while the upper sets represent the data de-reddened using two different reddening laws. Both reddening laws assume log(g)=4.0 and a distance of 121pc. The magenta curve is an A0 stellar template dereddened according to \citet{Weingartner:2001} Case B with R$_{V}$=5.5 and A$_{V}$=11.5, which yields L=14.3L$_{\sun}$. This was the assumption of \citet{Brown:2012}. The blue line is an A0 template dereddened according to \citet{McClure:2009}, which may be more appropriate for this high extinction region. This reddening assumption yields a factor of $\sim$1.7 higher dereddened luminosity of 23.6L$_{\sun}$. The upper curve tracks the stellar photosphere more closely out through the NIR, calling the presence of a hot dense inner disk component into question}
\label{fig:red_comp}
\end{figure*}

\begin{table}[b]
\caption{Previously published photometry and spectra of Oph IRS 48.}
\centering
\begin{tabular}{ c c c }
\hline\hline
Wavelength(s) & Instrument & Reference \\
($\mu$) & & \\
\hline
0.43, 0.64 & NOMAD & \citet{Zacharias:2005} \\
0.65, 0.8 & Hydra & \citet{Erickson:2011} \\
1.2, 1.6, 2.2 & 2MASS & \citet{Cutri:2003} \\
3.4, 4.6, 12, 22 & WISE & \citet{Wright:2010} \\
3.6, 4.5, 5.8 & Spitzer IRAC & \citet{van-Kempen:2009} \\
5.9-36.9 & Spitzer IRS & \citet{McClure:2010} \\
18.7 & AKARI IRC & \citet{Yamamura:2010} \\
60-181 & Herschel PACS & \citet{Fedele:2013} \\
70 & Spitzer MIPS & \citet{van-Kempen:2009} \\
450, 850, 1300 & SCUBA & \citet{Andrews:2007} \\
\hline
\end{tabular}
\label{table:SED}
\end{table}

\subsection{Oph IRS 48 Age and Spectral Type}

In order to derive the intrinsic unreddened properties of IRS 48 we first investigated the spectral classification using the SpeX spectrum of the star. We compared the spectrum directly to similar data on other stars. The spectrum of IRS 48 exhibited noticeable line emission within the photospheric features at wavelengths longer than 1.0 \micron{} (Pa $\gamma$, Pa $\beta$, etc.) so only photospheric lines at shorter wavelength were employed (Pa $\delta$, Pa $\epsilon$, Pa 09, etc.). The lines were found to be a good match to SAO 206463 (A0V, shown overplotted in Figure \ref{fig:SpeX-a}) so we adopted a spectral type of A0$\pm$1. This is in agreement with the spectral type determined by \citet{Brown:2012a}. 

To place Oph IRS 48 on a pre-main sequence evolutionary track, the absolute luminosity needs to be determined by dereddening the data. We investigated dereddening the spectrum using two reddening laws, that of \citet{Weingartner:2001} and that of \citet{McClure:2009}. For the former, we used their Case B reddening law, with $R_v$=5.5. In order to match the spectral slope of a A0 star, an extincion of $A_V$=11.5 mag was required. This was the methodology used by \citet{Brown:2012} in obtaining their luminosity estimate of 14.3 L$_{\sun}$. 

However, \citet{McClure:2009} found that for $A_{K}>$1 mag in dark clouds, a much``greyer" extintion curve applies, with $R_V$ closer to 8.5. Using this extinction law, we derive $A_V$=12.0 mag (determined independently from, but consistent with, the \citet{McClure:2009} estimate of $A_V$=12.9 mag for this object) and $A_K$=1.7 mag. 

The greater total reddening correction using this extinction law requires a larger stellar radius and luminosity, the latter increasing by a factor of $\sim$2 compared to that retrieved  by \citet{Brown:2012} using the \citet{Weingartner:2001} Case B extinction curve, to 23.6L$_{\sun}$. Hence, there is a large intrinsic uncertainty in the stellar luminosity derived, through the choice of extinction law adopted.

A comparison between the two dereddened spectra is shown in Figure \ref{fig:red_comp}. While both provide reasonable fits to the existing photometry and spectra, the \citet{McClure:2009} fit follows the stellar photosphere much more closely out through the NIR. This calls into question the existence of a hot inner disk component in this object, which has been inferred from the presence of excess emission at these wavelengths \citep[e.g.][]{Bruderer:2014,Maaskant:2013}. 

We note that neither reddening law provides a particularly good fit to the available optical photometry \citep{Zacharias:2005, Erickson:2011}. Given the inconsistency between the two published R-band magnitudes, we believe that additional photometry or spectroscopy at visible wavelengths is advisable in order to further constrain the best choice of reddening law for this object. 

\begin{figure}
\includegraphics[scale=0.35]{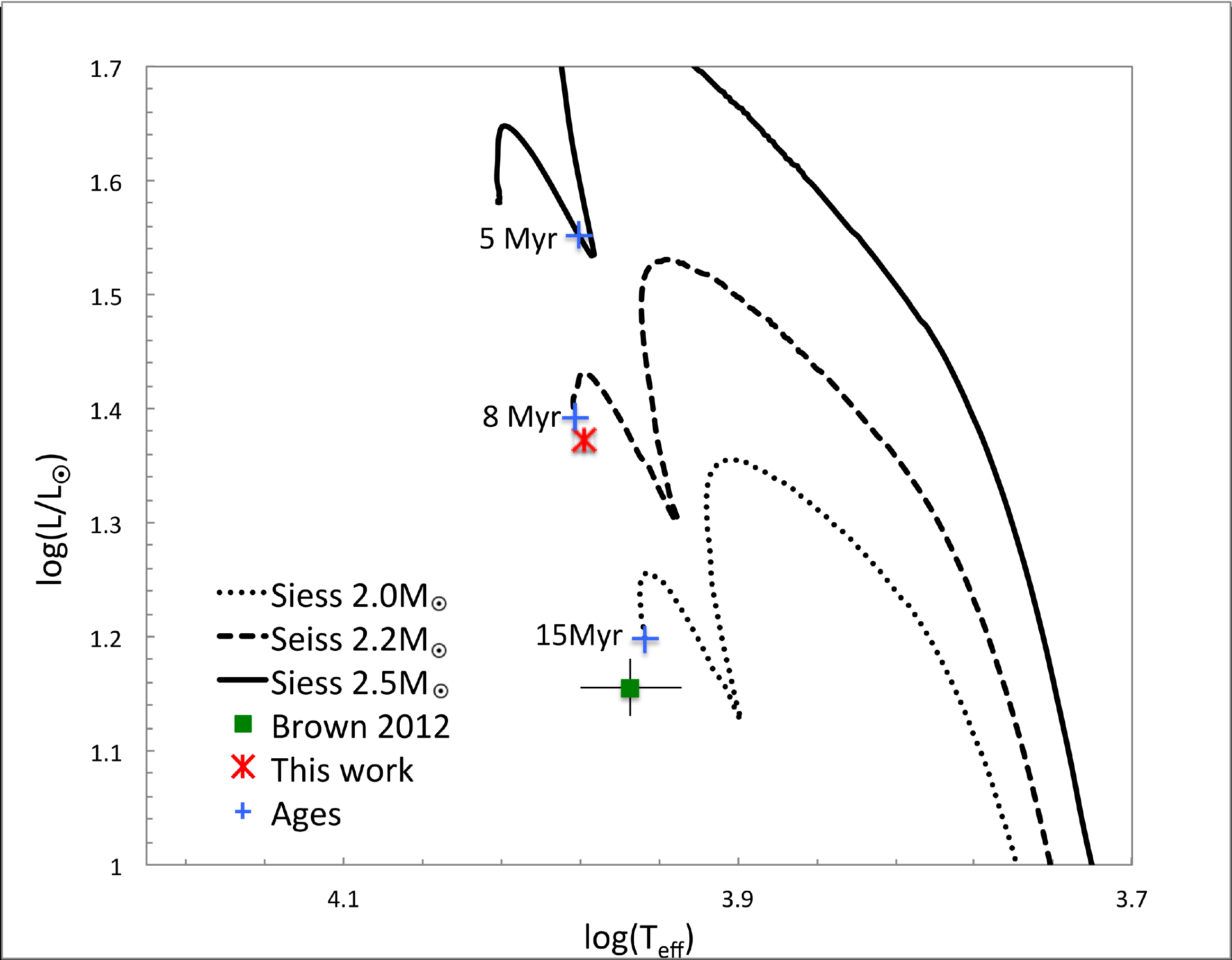}
\caption{An H-R Diagram showing the location of Oph IRS along \citet{Siess:2000} PMS evolutionary tracks. The green square shows the location of Oph IRS 48 according to the \citet{Brown:2012} reddening assumption, while the red asterisk shows the location of Oph IRS 48 derived from our best-fit dereddening. The factor of $\sim$1.7 difference in the luminosity derived from the two reddening assumptions results in different assumptions about the mass and age of Oph IRS 48. \citet{Brown:2012}'s \citet{Weingartner:2001} law puts Oph IRS 48 at $\sim$15Myr along the 2.0M$_{\sun}$ evolutionary track. We find reasonably well-fitting SEDs with Luminosities as high as 45L$_{\sun}$, corresponding to a 2.5M$_{\sun}$ star at 5Myr, and our best fit \citet{McClure:2009} law puts it at $\sim$8Myr along the 2.2M$_{\sun}$ evolutionary track. Whether or not these isochrones are the most appropriate for young A-stars is a matter of contention, however our higher derived luminosity puts Oph IRS48 along consistently higher mass isochrones and at a younger age than previous work regardless of chosen evolutionary model, as described in the text.}
\label{fig:HRdiag}
\end{figure}

These two different reddening laws also put the star at different places along pre main sequence evolutionary tracks. The \citet{Weingartner:2001} reddening law adopted by \citet{Brown:2012} puts the star at an age of $\sim$15Myr along the 2.0 M$_{\sun}$ PMS evolutionary track of \citet{Siess:2000}. Our factor of 1.7 higher luminosity using the \citet{McClure:2009} law puts the star at an age of $\sim$8Myr along a 2.2M$_{\sun}$ evolutionary track. 

We can derive well-fitting SEDs up to significantly higher stellar luminosities, corresponding to masses up to 2.5M$_{\sun}$ and ages as young as 5 Myr, this the error in our (and other) luminosity determinations is quite large. We note that although Pre-Main Sequence evolutionary models do not agree well in general, especially among higher mass stars \citep[e.g.][]{Palla:1993, DAntona:1994, Siess:2000}, this difference in inferred age as a result of reddening assumptions is largely model independent. 

Although we have used \citet{Siess:2000} here to match \citet{Brown:2012}, the age and mass inferred for Oph IRS 48 should be systematically younger and higher respectively for all evolutionary models under this \citet{McClure:2009} reddening assumption due to the higher inferred luminosity of the star. If we use our derived luminosity and measured temperature and compare them to other available pre-main sequence isochrones for A stars \citep[e.g.][]{Palla:1999, Dotter:2008}, our age estimate can fall to just a few Myr. 

Age estimates for other rho-Ophiuchus members within 10 arcminutes of Oph IRS48 range from $<$0.1Myr to 14Myr \citep{Wilking:2005}, so cannot place any meaningful constraints on evolutionary models, although the younger age that we derive using a \citet{Siess:2000} evolutionary track is more closely consistent with the median age of 2-5 Myr for stars in the region \citep{Wilking:2008}.

\subsection{Accretion Rate Estimate}
We were also able to estimate the stellar accretion rate from the SpeX data by fitting the Pa$\beta$ and Br$\gamma$ lines, and a sample fit is shown in \ref{fig:SpeX-b}. To do this, we subtracted our best fit stellar template (A0) from the dereddened SpeX spectrum to get the accretion line luminosities. Following \citet{Muzerolle:1998} for both lines and \citet{Calvet:2004} for an independent estimate of Br$\gamma$, we derive log(L$_{acc}$)=-1.1 for Pa$\beta$ and log(L$_{acc}$)=-1.0 for Br$\gamma$, which translate to $\dot{M}$=10$^{-8.5}$ and 10$^{-8.4}$ M$_{\sun}$/yr respectively using the relationship:
 \begin{equation}
\dot{M}=\frac{L_{acc}R_{*}}{0.86GM_{*}}
\end{equation}  
Our best estimates for the accretion luminosity and mass accretion rate are thus log(L$_{acc}$)=-1.1 and $\dot{M}$=10$^{-8.5}$ M$_{\sun}$/yr for Oph IRS 48. These are slightly lower than the log(L$_{acc}$)=-0.9 and $\dot{M}$=10$^{-8.4}$ derived in \citet{Salyk:2013} using the Pfund$\beta$ line at 4.7$\mu$m, but well within the $\sim$0.5dex spread in their empirical relationship.

\section{Discussion} 

\subsection{Monochromatic Radiative Transfer Modeling wth Sprout}
\label{sprout}
We began our modeling efforts by performing monochromatic radiative transfer calculations using a Monte Carlo code developed for use with SEEDS observations \citep[the Sprout code\footnote{Available at http://www.asiaa.sinica.edu.tw/\~{}jkarr/Sprout/sprout.html}, ][]{Takami:2013} in order to constrain disk parameters, particularly flaring and scale height, for full MCRT modeling. With Sprout, the modeled system consists of an illumination source with an axisymmetric circumstellar disk following a standard flared accretion disk prescription \citep[e.g.,][]{Shakura:1973,Lynden-Bell:1974} and described in cylindrical coordinates ($r$,$z$) by:
\begin{equation}
\rho (r,z) = \rho_0 \left[1-\sqrt{\frac{R_*}{r}} \right] \left(\frac{R_*}{r} \right)^\alpha ~ \rm exp~ \left\{- \frac{1}{2} \left[\frac{\it z}{\it h}\rm \right]^2 \right\},
\end{equation}
where $\rho_0$ is a constant to scale the density, $R_*$ is the stellar radius, $\alpha$ is the radial density exponent, and $h$ is the disk scale height. The scale height $h$ increases with radius as $h = h_0 (r/r_0)^\beta$, where $\beta$ is the flaring index ($\beta > 0$) and $h_0$ is the scale height value at a specific radius $r_0$, which we fix at 60AU to match our data. This schema is commonly used by other researchers \citep[e.g.,][]{Cotera:2001,Whitney:2003,Robitaille:2006,Follette:2013,Takami:2013}. 

We also use the common assumption $\alpha = \beta+1$ \citep[e.g.,][]{Robitaille:2006,Robitaille:2007,Follette:2013,Takami:2013}.This provides a surface density of $\Sigma \propto r^{-1}$, the same prescription used by \citet{Bruderer:2014}. For these simulations, we use the dust model developed by \citet{Cotera:2001} to explain the scattered light in the HH 30 disk. 

The free parameters for the density distribution of the disk are  therefore $\rho_0$, $h_0$, $\beta$. We assume that the mass of small dust, which is responsible for scattered light from the disk surface is $2.4 \times 10^{-6}$M$_{\sun}$, 15$\%$ of the total (large + small grain) dust mass derived in \citet{Bruderer:2014}. For a given $\beta$, we adjust the disk scale height $h_0$ to match the Northward offset of the observed $PI$ distribution from the star, and $\rho_0$ to achieve the above dust mass. We find that the flaring parameter $\beta$ primarily affects the brightness at outer radii compared with that at inner radii and that we generally need values of $\beta>$1 to reproduce our data.

\begin{figure*}
\begin{center}
\includegraphics[scale=0.75]{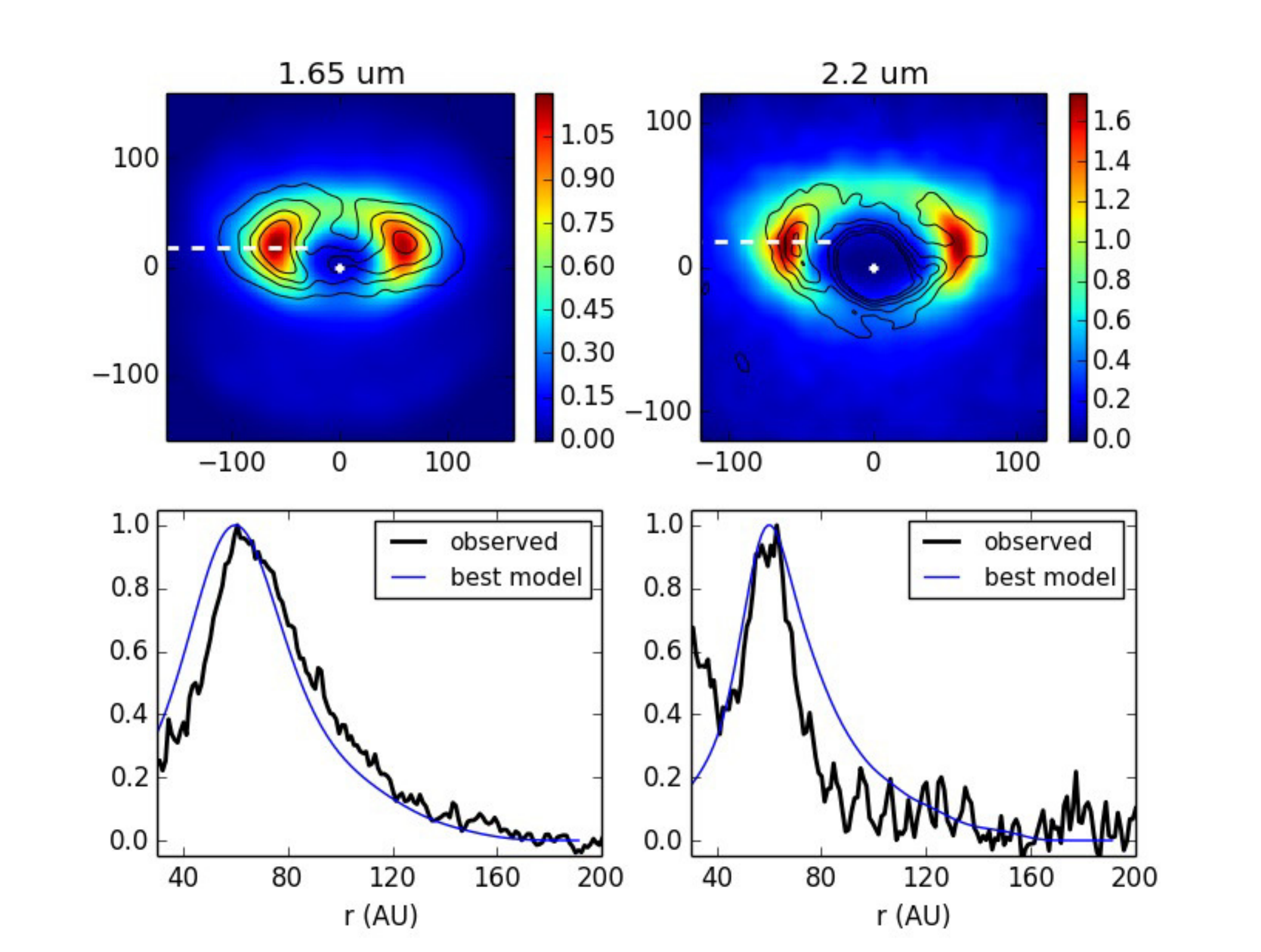}
\end{center}
\caption{Top: Simulated images of Oph IRS48 at H-band (left) and Ks-band (right) generated by the monochromatic radiative transfer Sprout code. The data are shown overplotted in each case as black contours. The location of the illumination source is marked with a white diamond, and the locations where profiles were extracted are marked with white dashed lines, corresponding roughly to the ''major axis" relative to the apparent cavity center. The combination of an elevated flaring parameter ($\beta$=1.3) and a larger scale height at the cavity rim (15AU) are able to reproduce the apparent Northward offset of the scattered light cavity from the star, verifying that it is likely a result of viewing geometry and is not a true offset, as described in detail in the text. Bottom: Simulated radial polarized intensity (blue) profiles for these models shown relative to the data at H (left) and Ks (right).}
\label{fig:Takami}
\end{figure*}

We used $10^6$ photons for each simulation, collect the scattered light at viewing angles of $50^\circ \pm 10^\circ$, and convolve the simulated images with a Gaussian whose FWHM matches the observations. Photons that do not interact with the disk are not collected. Figure \ref{fig:Takami} shows one of the best fit models ($\beta$=1.3, $h_0$=15AU at $r$=60 AU). 

This $\beta$ is similar to the model by \citet{Bruderer:2014} used to explain images at millimeter and mid-infrared wavelengths (1.2), and is within the typical range for a flared disk (to first approximation, temperature varying as r$^{-1/2}$ in a disk typically leads to a scale height proportional to r$^{5/4}$), so we adopt it for all further modeling efforts. 

The scale height required in the Sprout model, however, is $\sim$2x the \citet{Bruderer:2014} small grain model (8.4 AU), corresponding to a midplane temperature difference of 4x, which is not explicable simply by the greater stellar luminosity derived in this work (our factor of 1.7 increase in luminosity corresponds to just a factor of 1.2 in temperature at a given location). As this large scale height is required to reproduce the observed Northward offset, we utilize large gap edge scale heights in our future modeling efforts, though we were somewhat limited in our ability to do so given the constraints of the models,  as described in  detail in Section \ref{offset}. 
 
\begin{figure*}
\includegraphics[scale=1.0]{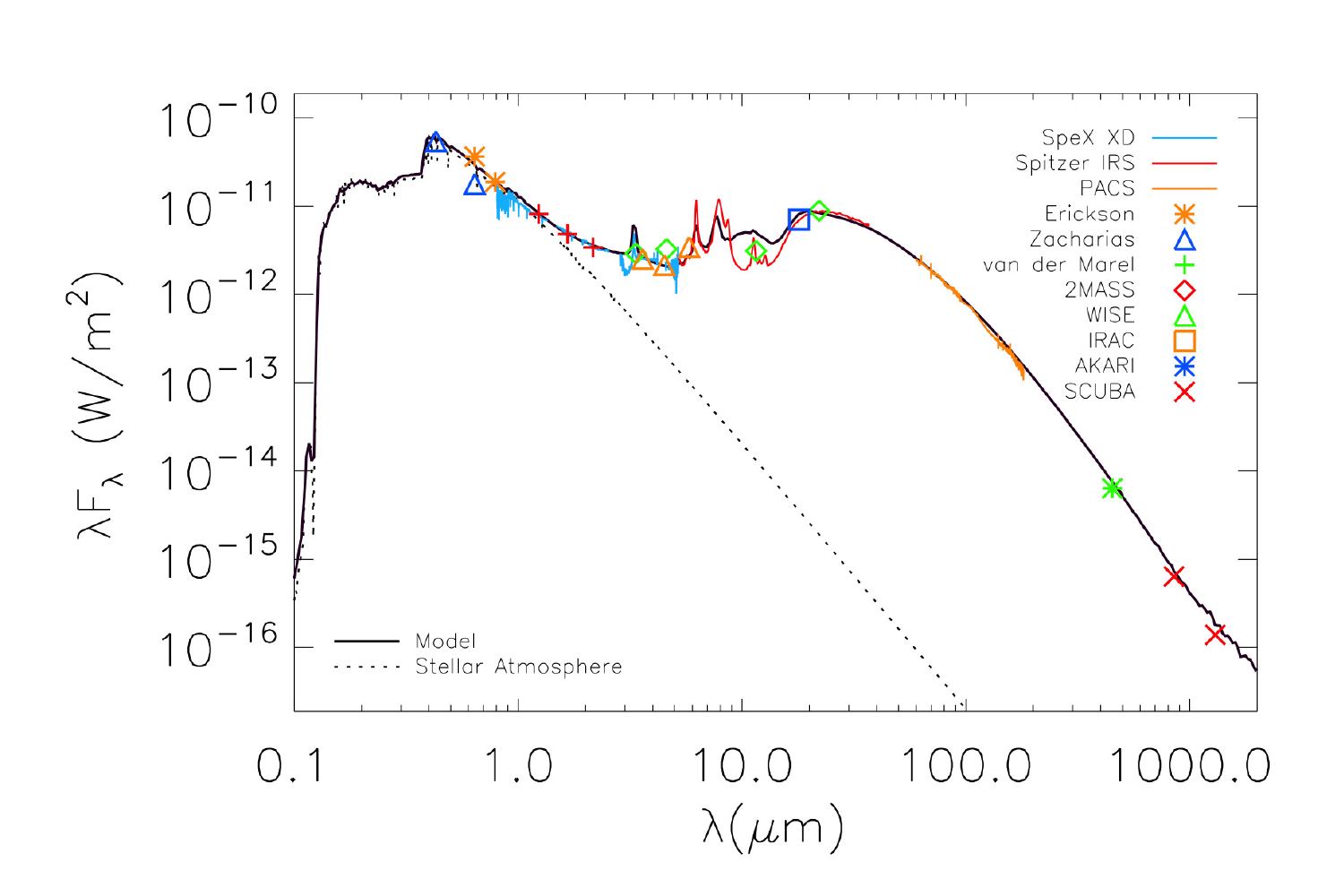}
\caption{The best fitting SED model generated with the Whitney code, shown in black and following the parameters outlined in Table \ref{table:modelpars}, overplotted on observational data from the literature. The sources for the literature data are described in detail in Table \ref{table:SED}. The poor quality of the fit in the 10$\mu$m region is discussed in Appendix \ref{10micron}.}
\label{fig:sed}
\end{figure*}

\begin{table}
\caption{}
\centering
\begin{tabular}{ c c c }
\hline\hline
Parameter & Value & Reference \\
\hline
M$_{star}$ & 2.2M$_{\sun}$ & this work \\
T$_{star}$ & 9500K & this work \\
R$_{star}$ & 1.8R$_{\sun}$ & this work \\
M$_{disk,dust}$ & 1.6$\times$10$^{-5}$ & \citet{Bruderer:2014} \\
R$_{max}$ & 160AU & \citet{Bruderer:2014}, this work\\
R$_{min}$ & 1AU & this work\\
R$_{gap}$ & 60AU & this work \\
$i$ & 50$^{\circ}$ & \citet{van-der-Marel:2013} \\
$\beta$ & 1.3 & this work \\
$\delta_{gap}$ & 0.004 & this work \\
$H_{60AU}$ & 10AU & this work \\

\hline
\end{tabular}
\label{table:modelpars}
\end{table}

\subsection{SED modeling}
Informed by the results of our Sprout simulations, we ran a suite of disk models using the updated 2013 version of the three dimensional Whitney monte carlo radiative transfer (MCRT) code \citep[see][for details]{Whitney:2013,Whitney:2003A,Whitney:2003}. We began our modeling investigation by attempting to fit the multiwavelength observed SED of Oph IRS 48, which includes a wide range of broadband photometry as well as optical, NIR and MIR spectra. Sources for this data are listed in Table \ref{table:SED}, and our best fit model is overplotted on data from the literature in Figure \ref{fig:sed}. This best fit SED is an excellent match to the literature photometry and spectra except at the 10$\mu$m silicate feature. Attempts to fit this portion of the SED, and their incompatibility with scattered light images generated by the models, is described in detail in Appendix \ref{10micron}.

Given the wide-ranging degeneracies of SED modeling, input parameters were fixed to observed values wherever possible, and only parameters having a marked effect on the quality of the SED fit were investigated in detail. Relevant parameters are listed in Table \ref{table:modelpars}. 

Whereas previous modelers have assumed a lower intrinsic stellar luminosity, and this has required an additional inner dust disk component \citep[e.g.][]{Bruderer:2014} or halo \citep[e.g.][]{Maaskant:2013} to reproduce the NIR excess seen in the SED, our choice of the \citet{McClure:2009} extinction law increases the intrinsic luminosity of the star and the dereddened stellar photosphere tracks the observed photometry closely out through the NIR. This contrast in the fit to the NIR photometry according to reddening assumption can be seen clearly in Figure \ref{fig:sed}. We adopt a simple uniformly depleted dust gap that extends inward to the sublimation radius in our dust modeling. The surface density profile of our model disk therefore closely follows that used by \citet{Andrews:2011}, and in particular their Figure 2a.

In all models, we assumed a simple disk structure with two components (large and small grains, with large grains restricted to 20$\%$ of the scale height of the small grains to mimic settling of large grains toward the midplane) and a disk rim at 60AU, inside of which the disk has the same general properties, but is depleted by a factor of $\delta_{gap}$. In our best fit model, $\delta_{gap}$=0.004, which is within a factor of a few of $\delta_{gap,dust}$=0.0009 inferred by \citet{Bruderer:2014}, although their dust structure is somewhat more complex, being wholly depleted from 1-60AU with an inner $<$1AU dust disk depleted by this factor. 

The \citet{McClure:2010} 6-37$\mu$m Spitzer IRS spectrum, which traces disk emission through the PAH bands in the NIR-MIR and across the thermal emission bump due to the outer disk rim places significant constraints on dust properties. As noted in \citet{Geers:2007}, the PAH features in this disk are very strong, pointing to the presence of gas inside of the cavity, thus it cannot be wholly depleted of material. Additionally, \citet{Geers:2007} were able to isolate H$\alpha$ emission in this object, providing evidence of ongoing accretion and further lowering the feasibility of a fully depleted central cavity. Our inability to fully remove the stellar residual through halo correction is further anecdotal evidence in favor of material inside the cavity. 

Treating the dust mass of 1.6$\times$10$^{-5}$M$_{\sun}$ derived in \citet{Bruderer:2014} from an observed gas mass of 1.5$\times$10$^{-4}$M$_{\sun}$ (and an assumed 10:1 gas:dust mass ratio) as a fixed parameter, we find that among the large grain dust prescriptions that are standard in the Whitney model, a steep power law grain size distribution (p=3.5) and large maximum grain size (1mm) is best able to reproduce both the \citet{Fedele:2013} 55-210$\mu$m Herschel PACS spectrum and the \citet{Andrews:2007} 850$\mu$m and 1.3mm SCUBA photometry.  A similar model with a smaller maximum grain size (20$\mu$m) drastically overproduces sub-mm flux unless the dust mass is set to a factor of 3 lower than the \citet{Bruderer:2014} value, and this model is also unable to reproduce the correct PACS slope. It is worth noting, however, that the assumption that grain sizes and compositions are uniform throughout the disk is certainly an oversimplification, as discussed in detail in \citet{Menu:2014}.

With these dust prescriptions and a dust mass of 1.6$\times$10$^{-5}$M$_{\sun}$ , we find that we are best able to match the SED by placing approximately 85$\%$ of the dust mass in large grains, and the remaining 15$\%$ in small ISM-like sub-micron grains. More specifically, we find that this 85/15$\%$ division of dust mass into large and small grains respectively is best able to reproduce the size and shape of the MIR spectral bump as well as the sub-mm photometry. This is similar to values found by other modelers for both this and other transitional disks \citep[e.g.][]{Andrews:2011}. 

\begin{figure}
\centering
\subfloat[][]{
\label{fig:modelims-a}
\includegraphics[scale=0.23]{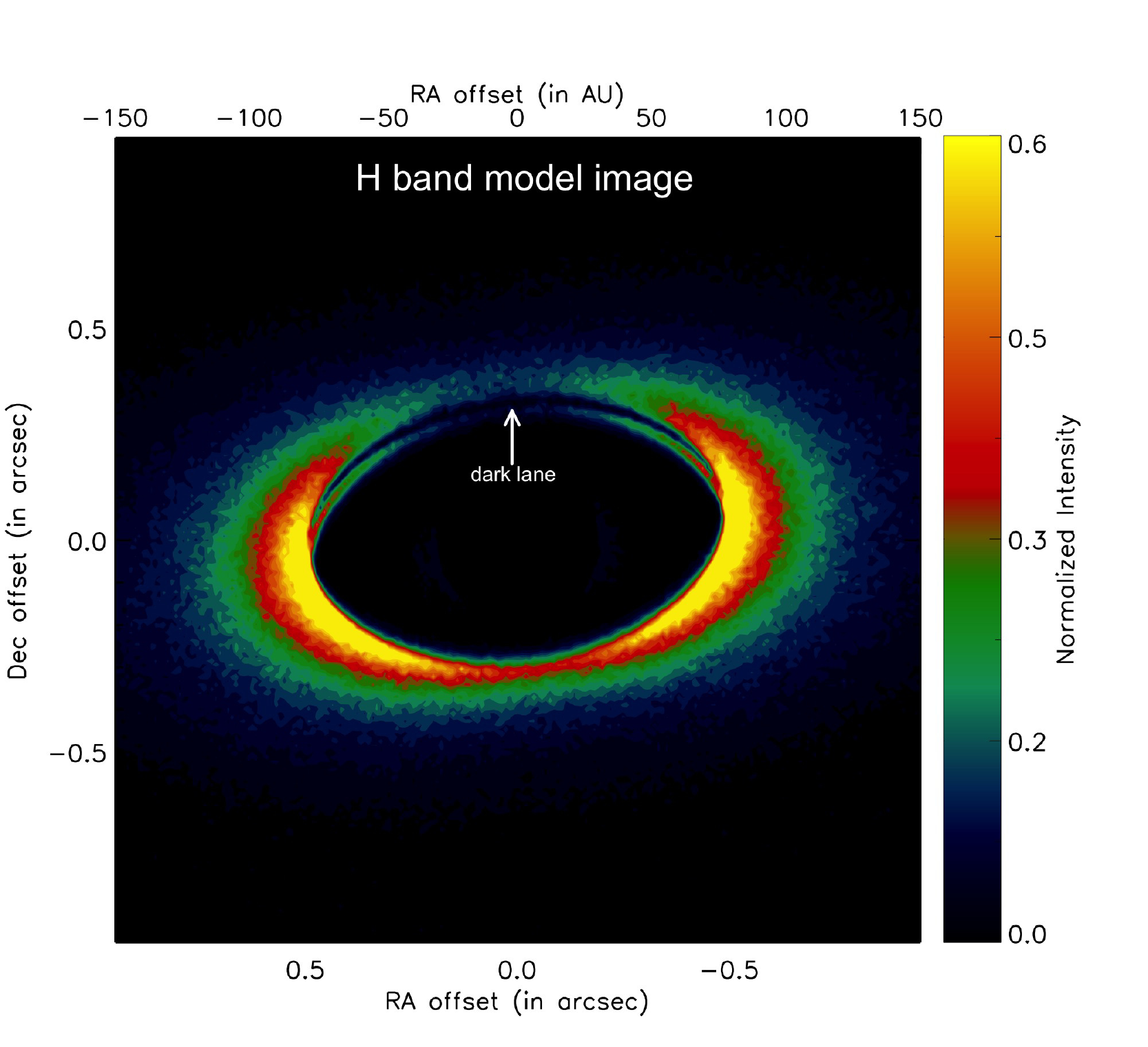}}
\hspace{8pt}
\subfloat[][]{
\label{fig:modelims-b}
\includegraphics[scale=0.23]{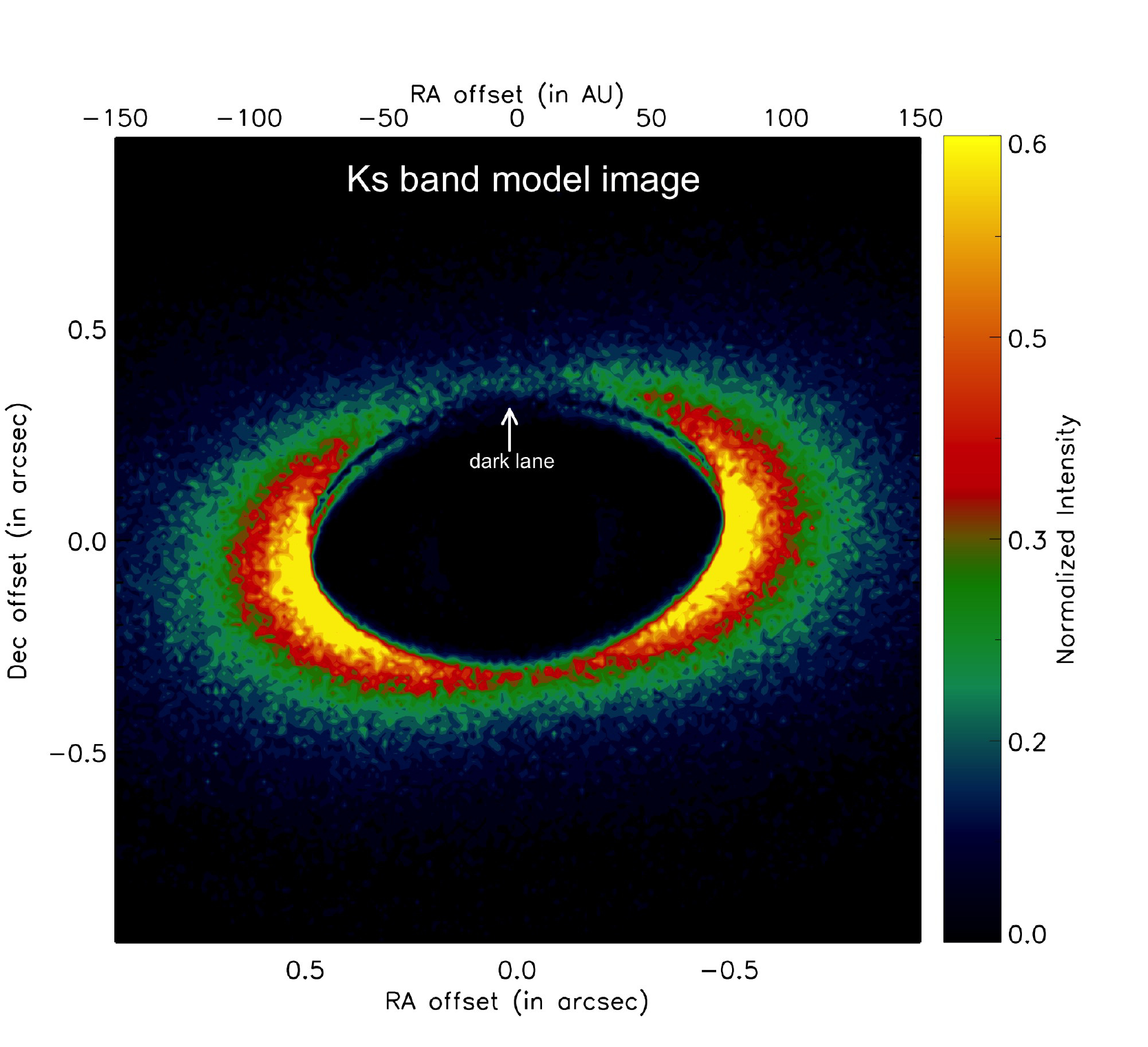}}\\
\subfloat[][]{
\label{fig:modelims-c}
\includegraphics[scale=0.23]{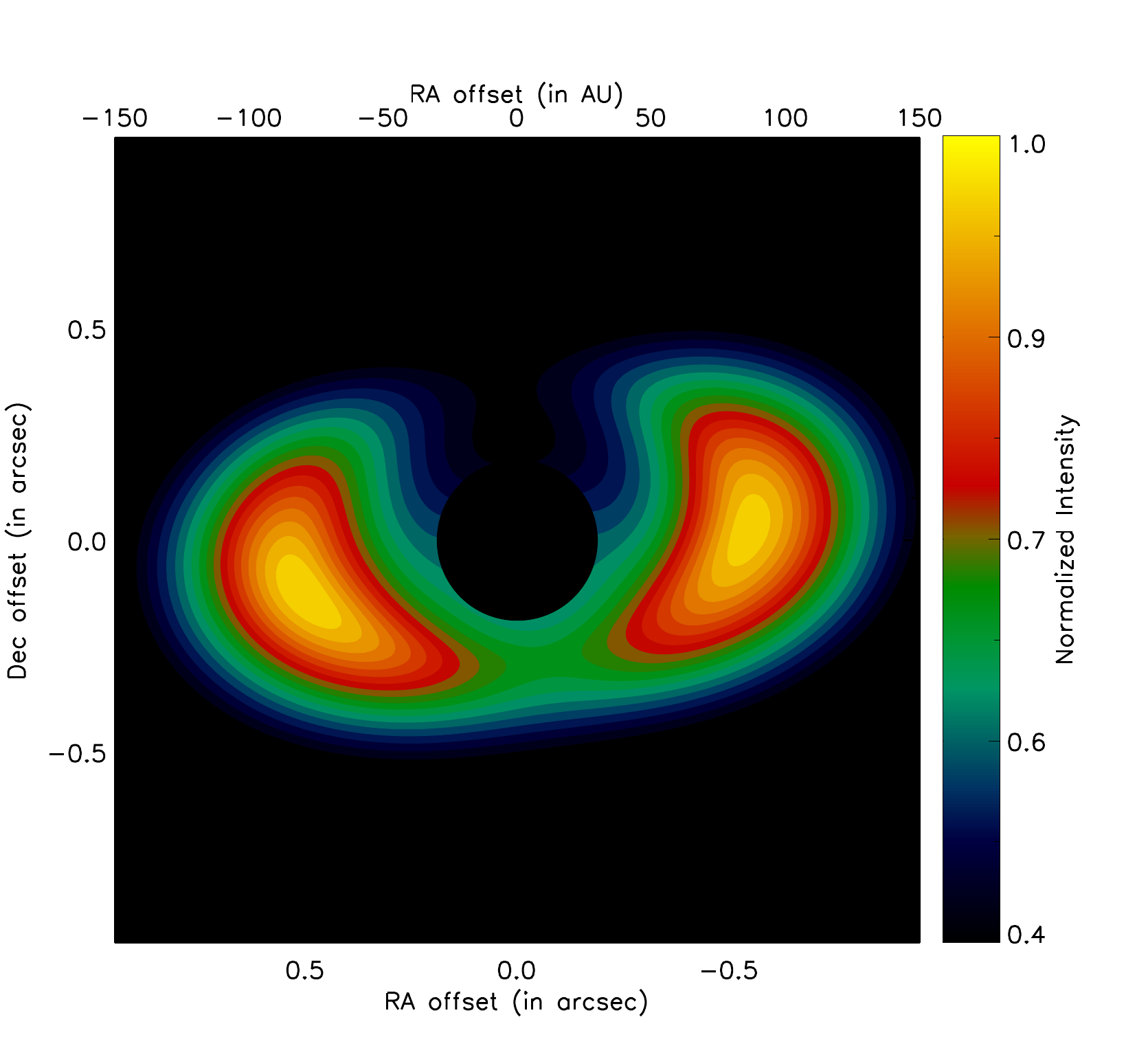}}
\hspace{8pt}
\subfloat[][]{
\label{fig:modelims-d}
\includegraphics[scale=0.23]{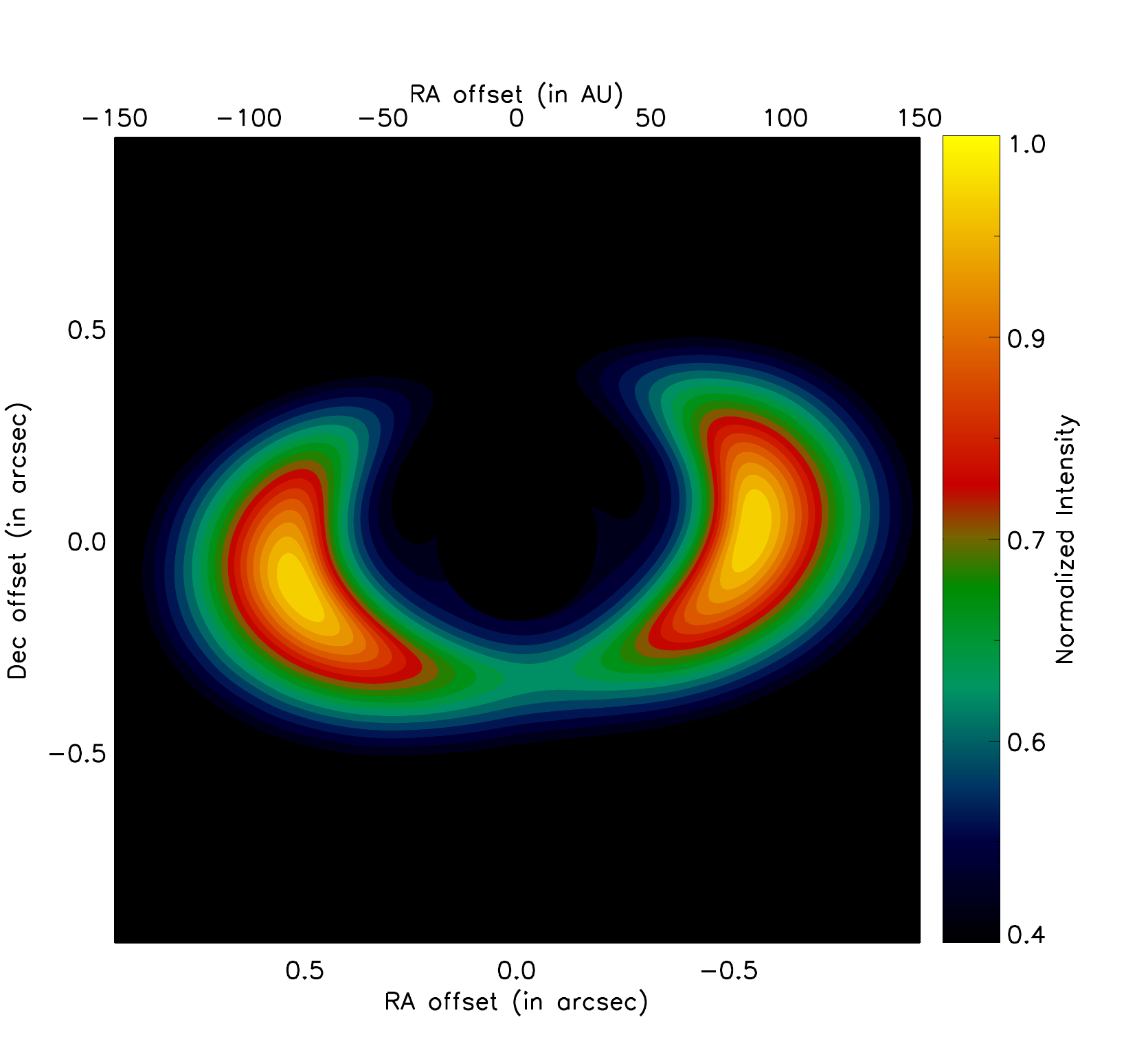}}\\
\caption{Simulated images of the Oph IRS 48 disk created with the Whitney code at H band (left) and K band (right). In the bottom panel, these images have been convolved with the observed PSF at each wavelength.}
\label{fig:modelims}
\end{figure}

\subsection{Modeling of Scattered Light Images}
\label{diskmodel}
Having narrowed the available parameter space significantly through SED modeling, we also attempted scattered light modeling of the disk with the Whitney code. We reran the Whitney code for our best-fitting SED model with a greater number of photons, as needed for full scattered light image simulations. We iterated on the output until we got a reasonable match to the geometry and brightness distribution of our observed images while maintaining the quality of the SED fit. Although this modeling effort was by no means exhaustive, it does inform some of the likely disk properties. 


The presence of a dark lane to the North in the model images shown in Figure \ref{fig:modelims} suggests that grains inside of the disk cavity cause some shadowing of the inner cavity wall to the North. Because of the geometry of the disk's inclination, the dark lane that presumably exists along the Southern rim is not visible.  This dark lane is apparent only in the raw model images, and is not evident once they are convolved with the PSF (bottom panels). We believe, therefore, that though a dark lane is not visible in our data, shadowing by the non-negligible amount of material within the cavity contributes to the faintness of the real disk images to the North. Higher resolution imagery of Oph IRS 48 in the future may serve to verify this conclusion. 

\begin{figure*}
\centering
\subfloat[][]{
\label{fig:modprofs-a}
\includegraphics[scale=0.45]{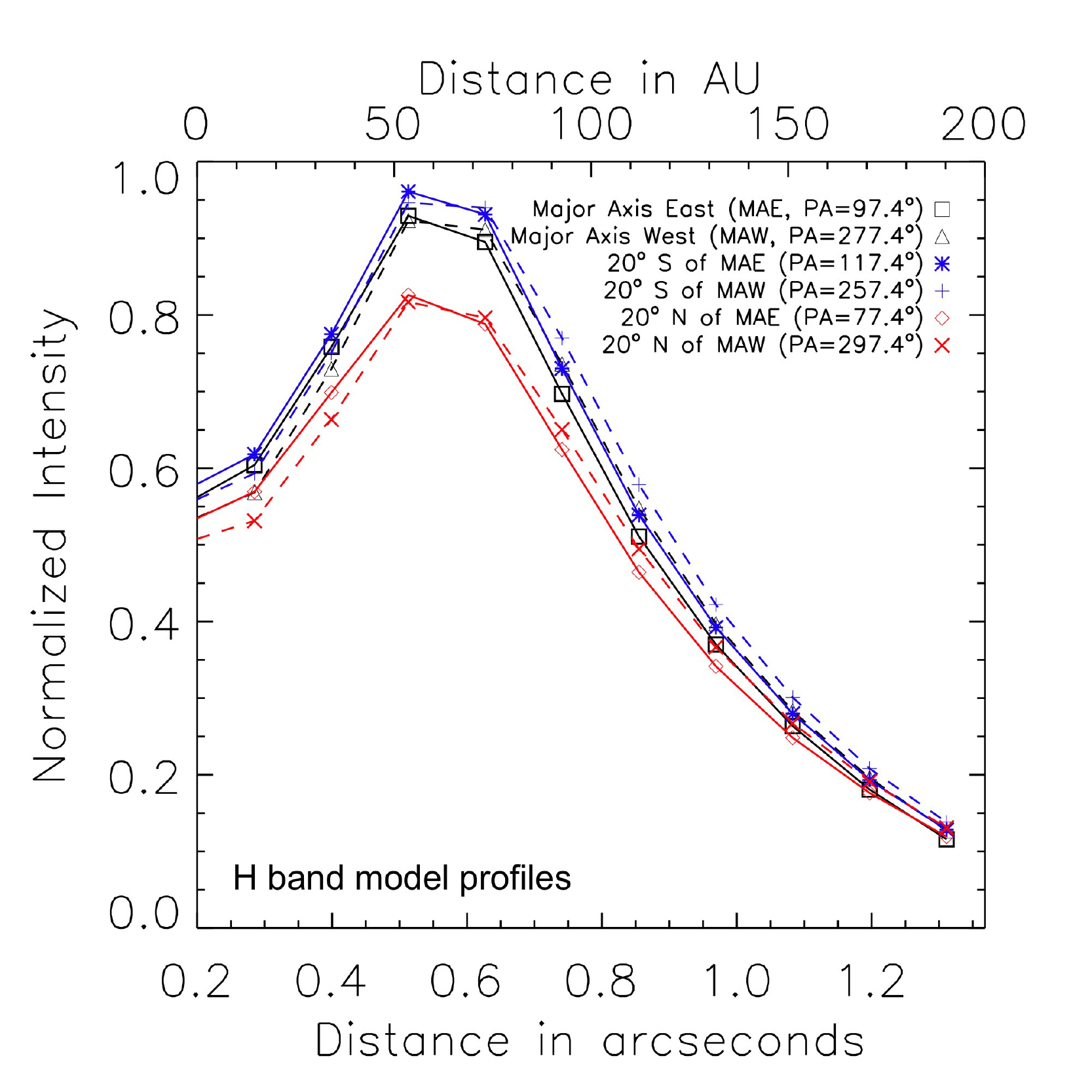}}
\subfloat[][]{
\label{fig:modprofs-b}
\includegraphics[scale=0.45]{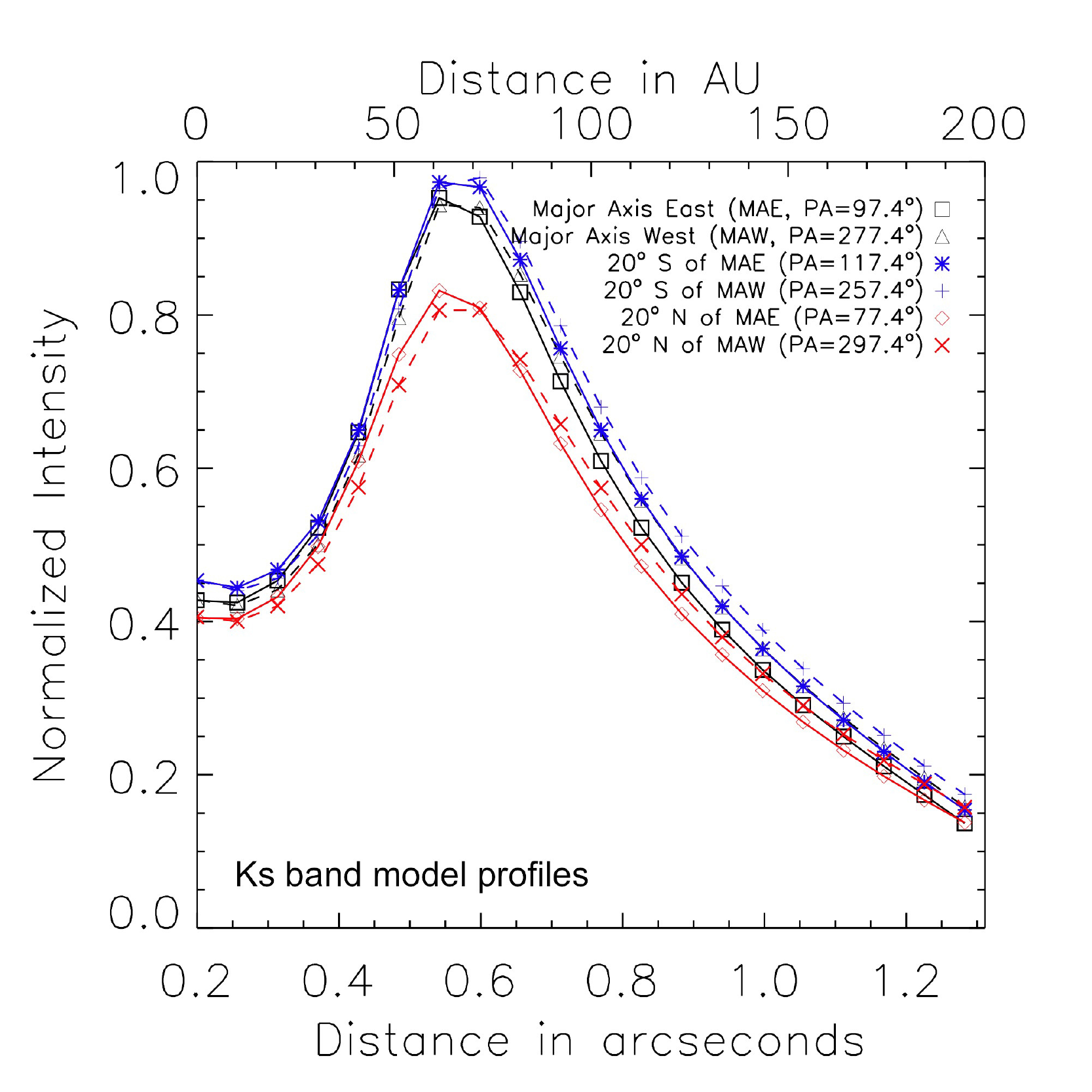}}\\
\caption[]{Radial profiles taken through the deprojected H (left) and K (right) model images, which are informative in contrast to the true observed radial profiles shown in Figure \ref{fig:rps}. The model profiles are drastically different from the observations in their East/West symmetry. They do, however provide a reasonable reproduction of the relative shape, location and relative intensity of the peaks outside of the spiral arm.}
\label{fig:modprofs}
\end{figure*}

Radial profiles were also taken through these model images and are shown in Figure \ref{fig:modprofs} and described in detail in Section \ref{asymm}.

The primary way in which modeled disk images appear to fall short of the reality of Oph IRS 48 is in their assumption of azimuthal symmetry. In particular, the Southwestern and Western parts of the disk, which host the ALMA excess and spiral arm respectively are significantly less uniform in brightness and smooth in morphology than both the disk models and the Eastern rim would suggest. Non-axisymmetric modeling is beyond the scope of this work.

\begin{figure}
\begin{center}
\includegraphics[scale=.8, trim={0.25cm 0 0 0},clip]{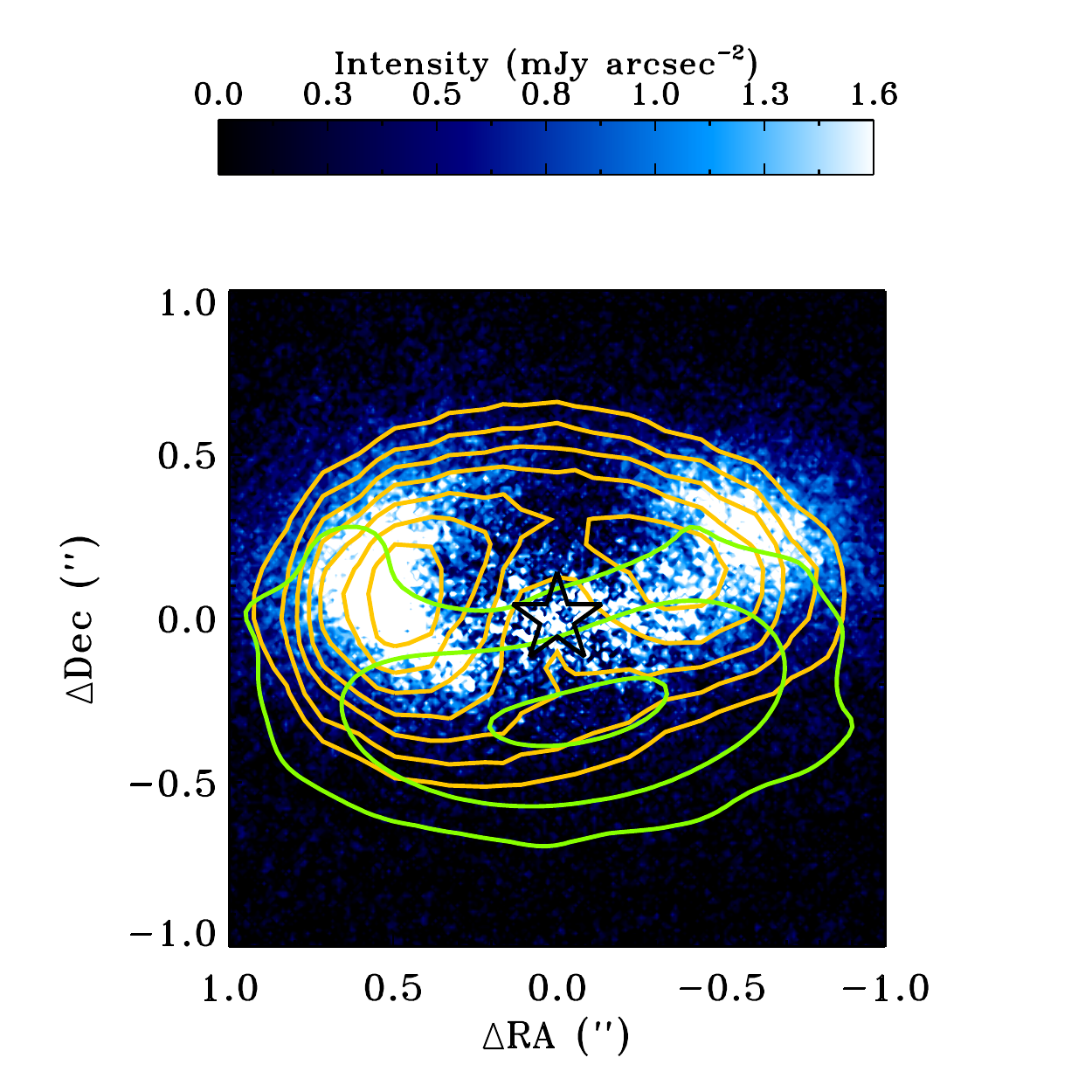}
\end{center}
\caption{Contour map of \citet{Geers:2007} 18$\mu$m VLT VISIR data in yellow and \citet{van-der-Marel:2013} 0.44mm continuum ALMA data in green overplotted on our HiCIAO median halo-subtracted H band PI image. Note that the 18$\mu$m thermal emission is centered on the star, and that its Eastern lobe is cospatial with ours, whereas the 18$\mu$m Western lobe lies inside of our scattered light lobe. As discussed in the text, this is evidence that the Western arc is fainter simply because it is farther from the central star}
\label{fig:overlay}
\end{figure}


\subsection{Brightness Asymmetries in Real and Modeled Data}
\label{asymm}
The disk of Oph IRS 48 as observed in the NIR contains a number of marked asymmetries. The most immediately apparent is the East/West brightness asymmetry, namely that the Eastern arc appears significantly brighter than the Western at both wavelengths. This is likely a simple result of proximity to the star, as the Western arc is as much as 0$\farcs$2 more distant from the central star, lying along a non-axisymmetric structure in the West of the disk that we discuss in detail in Section \ref{spiral}. 

This scenario would also explain the offset in the apparent position of the Western arc in our images relative to the 18.7$\mu$m contours of \citet{Geers:2007}, shown overlain on our PI image in Figure \ref{fig:overlay}. For a thermal mechanism, emission should be symmetric about the star, so we have assumed that the star is located at the center of the two lobes of 18$\mu$m emission. While this assumption places the Eastern lobe of the 18$\mu$m emission near the observed scattered light lobe, it places the Western H-band scattered light lobe almost entirely outside of its 18$\mu$m counterpart.


The scattered light models described in Section \ref{diskmodel} are able to reproduce the general shape and relative peak brightnesses of the Eastern disk profiles, although they deviate significantly in the West due to the model's incorrect assumption of azimuthal symmetry. We engage in more thorough discussion of the properties of the models in Sections \ref{offset} and \ref{spiral}, however we elected to stop our full MCRT simulations at this point, as we have obtained a reasonable facsimile of the general properties of the true disk and full MCRT simulations are computationally expensive.

\subsection{Offset of Cavity Center from Stellar Residual}
\label{offset}
As noted in Section \ref{cavmorph}, the center of the best fit ellipse to the Oph IRS 48 cavity in both data sets is offset from the stellar location. In the East/West direction, the offsets between the stellar residual and the cavity center are small, and are not uniform between the two datasets. In the Ks band data, the best fit cavity is offset by 5 pixels (0$\farcs$05) to the East of the star, and in H band it is offset by 6 pixels (0$\farcs$06) to the West. In both cases, this may be due to the influence of the spiral arm component on the cavity fit. In the H-band data, the asymmetry is not resolved and is therefore not excluded from the cavity fit, though it is clear from the overlay in Figure \ref{fig:HKoverlay} that the Western H-band arc lies mainly along the asymmetry, so this is likely to pull the best fit ellipse center to the West. In the case of the Ks-band data, the exclusion of the region of the asymmetry from the fits and the fact that the Eastern profiles are brighter means that the Eastern arc dominates the fit, and may pull the best fit ellipse center to the East. Given the small magnitude of these offsets and the fact that they are in opposite directions in the two data sets, we do not consider either one to be a true reflection of disk properties.

The large Northward offset of the best fit cavity center from the stellar residual at both wavelengths (15 pixels/0$\farcs$14 at H-band and 18 pixels/0$\farcs$17 at Ks band) is much larger and requires explanation. Our initial scattered light modeling of the disk, described in detail in Section \ref{sprout}, was able to reproduce a Northward offset of equal magnitude through simple disk geometry, namely a large scale height at the disk rim (Z$\sim$15AU). Since the Southern disk rim is shielded from our view, this tends to push the apparent Cavity center northward relative to our line of sight. 

If we take our best fit model images from the full MCRC modeling described in Section \ref{diskmodel} and use the same ellipse fitting procedure that we used in Section \ref{cavmorph}, then we get similar, though not identical, values for the best fit ellipse parameters. We derive a slightly larger (a=0$\farcs$58, b=$\farcs$42) cavity than using the true K-band data (a=0$\farcs$49, b=0$\farcs$37), though it is of similar eccentricity (0.69 vs. 0.66) and PA (96.3 vs. 97.2$^{\circ}$). We also reproduce a northward offset of the cavity center of between 6 and 9 pixels, depending on the exact parameters for the radial profile fits. This is about half of the observed offset. Our NIR modeling with the Sprout code suggests that simply increasing the scale height of the gap wall further should increase this offset to the observed value, however doing this with the Whitney model resulted in a larger MIR bump at $\sim$20$\mu$m than is observed in this object. We believe both the larger cavity radius and smaller scale height implied by the model are spurious, and are likely a result of the chosen grain prescription and disk geometry, however we leave a full exploration of the effect of these parameters on model images for future efforts. 

\subsection{On the Nature of the Western Extension}
\label{spiral}
The geometry of what we've called the Western arc of the Oph IRS48 disk, and in particular the nearly right angle at which it meets the Southern Ks-band arc, suggest that, at the very least, the disk of Oph IRS48 is not axisymmetric. Rather than tracing the cavity all the way around, emission seems to be concentrated at a significantly greater stellocentric distance along -100$<$PA$<$-70, explaining both its relative faintness in this region and the deviation of the location of peak brightness along these PAs from the cavity fit.  

One explanation for this deviation is that the emission in this region traces a spiral arm that extends to the Southwest from the Western cavity rim. ALMA gas data reveal that the Western half of the Oph IRS 48 disk is rotating counterclockwise away from Earth \citep{van-der-Marel:2014}. This means that the arm trails the disk rotation, as is observed in other disk spirals \citep{Grady:2013,Muto:2012}. By fitting the radial profile peaks in this region with a line in the deprojected cavity image and comparing the slope of this line to a circle at similar radius, we estimate the pitch of the arm to be $\sim$40$^{\circ}$. This value is quite high compared to values seen in other disk spirals \citep[e.g 15$^{\circ}$][]{Muto:2012}, indicating either a region of high temperature in the disk (H/R$\sim$1, consistent perhaps with the large scale height necessary to explain the northward offset of the cavity center from the stellar residual in our modeling) or that the portion of the arm that we are seeing is very close to the planet's corotation radius. 

Alternatively, this feature could be a local ``divot" in the disk rim at PA$\sim$270. Localized surface brightness deficits have been observed in other disks \citep[e.g.][]{Mayama:2012}. In terms of a physical explanation for such phenomena, the simulations of \citet{Jang-Condell:2009} suggest that a massive planet embedded in the disk could mimic this feature by creating a ``shadow" at the location of the planet (due to a rapid change in the disk scale height in this region), and a bright exterior region where the disk surface emerges from shadow. 

Regardless of the physical cause of this feature, H and K band profiles along the Western asymmetry, when compared to profiles through the rest of the disk, reveal that grain properties also deviate in this region.  A close examination of the deprojected radial profiles shown in Figure \ref{fig:rps}, reveals that the brightness of scattered light in this region relative to the rest of the disk varies according to wavelength. In particular, the profiles at 20$^{\circ}$ North of the disk major axis to the East and West (red lines) deviate between the H and K band data. When normalized to the Eastern profile peak, the Western profile is 25$\%$ brighter in the H-band data ($\sim$80-85$\%$) than in the Ks-band data ($\sim$55-60$\%$) even before the latter is smoothed to simulate similar AO correction, which reduces its brightness relative to the Eastern peak by an additional $\sim$5-10$\%$. 

Although the same numerical comparison can't be made along the major axis (black lines), since the cavity rim is not clear in the Western profile, the trend appears to hold here as well, and the Western profile remains bright out to a larger radius in the H-band data than in the Ks-band. This is not the case, however, for all other profiles, including all of the Southern (blue lines) and Eastern (solid lines) profiles, which are well matched between the smoothed Ks-band data and the H-band data. 

This effect is not present in the model profiles shown in Figure \ref{fig:modprofs}, which hints at a physical difference in the grains in this region relative to the rest of the disk. Specifically, the grains in the region of the asymmetry appear to have bluer scattering properties, which Mie theory would suggest is an indication that particles here are smaller than elsewhere in the disk. Dynamical perturbations may explain this deviation, as small particles should be well-coupled to the gas in the disk, while large particles have larger stopping times and may lag behind any gas perturbations \citep{Lyra:2009}. 

Although purely speculative, we note that this feature may somehow be associated with the dust enhancement observed by \citet{van-der-Marel:2013} and shown overplotted on our data in Figure \ref{fig:overlay}, which is located at a similar radius, though it lags behind the observed non-axisymmetric scattered light region by at least 30$^{\circ}$. 

Generally speaking, azimuthal asymmetries such as those observed in Oph IRS 48 are difficult to create through many of the alternative disk clearing mechanisms (photoevaporation, grain growth), and are most consistent with the existence of a perturber (planet or brown dwarf) in the disk. The gravitational influence of this body serves to both excite the observed asymmetries and clear the central cavity, perhaps with the assistance of other mechanisms. Some disk dynamical processes can create asymmetries, however these are either associated with gap opening (e.g. by a planet, Rossby-wave instability), require a lower gas-to-dust ratio than is observed in Oph IRS48 (photoelectric instability) are smaller in spatial scale than observed here (MRI-induced asymmetry) or are associated with immediate planet formation (gravitational instability).  Together with the \citet{Bruderer:2014} derivation of a two-step decrease in gas surface density inside of the cavity and the \citet{van-der-Marel:2013} observation of a pronounced sub-mm asymmetry in this disk, our data lend additional credence to the assumption that this disk hosts one or more planets.

\section{Conclusion}

We have spatially resolved the circumstellar disk of Oph IRS 48 for the first time at NIR wavelengths.  New H and Ks-band scattered light imagery reveal a cleared central cavity with a rim at $\sim$60 AU. Fits to the shape of the cavity are consistent with a circular geometry, but with a cavity that is offset to the North relative to the star due to viewing geometry.

The disk hosts a number of interesting asymmetries. The East is brighter than the West, explicable via the fact that the Western disk arc appears to be farther from the star than the Eastern arc. We interpret this feature as either (a) tracing a spiral arm with a pitch of $\sim$40$^{\circ}$ or (b) a local surface brightness defecit at PA=270$^{\circ}$ potentially caused by a ``planet shadow". 

The disk is a factor of $\sim$2 brighter in absolute intensity at Ks-band than at H-band, however this is likely a simple result of the highly extincted region in which Oph IRS 48 lies. The scattering properties in the region of the spiral arm, however, are markedly blue relative to the rest of the disk, suggesting that grains in this region are smaller than elsewhere in the disk. 

Scattered light modeling using compact grains is able to reproduce a reasonable facsimile of the true disk emission with two important caveats. First, the SED in the region of the 10$\mu$m silicate feature proves to be very difficult to match with most standard grain models. Secondly, the models are not equipped to reproduce azimuthal asymmetries such as are clearly present in the Western half of the Oph IRS 48 disk, and are only able to reproduce the relative shapes and brighnesses of radial brightness profiles along the Eastern disk arc. Neither of these problems is unique to this particular disk, and both suggest future directions for circumstellar disk models. 

Taken as a whole, these data on Oph IRS 48 serve to emphasize several key points that are true of high contrast disk imaging in general. First, the extinction assumption that is used to deredden photometry for modeling can have profound effects on conclusions about stellar properties. In our case, we find that our choice of reddening law drastically affects the luminosity and mass derived for the central star. The geometry of this disk at NIR and sub-mm wavelengths also highlights the complications of observed asymmetry in disks. In Oph IRS 48, an azimuthally symmetric disk cannot reproduce the observed data.

Our data are consistent with the existence of at least one planet in the disk, which could function both to clear the central cavity and to incite the observed deviation from axisymmetry. Oph IRS 48 is thus a good candidate for high-contrast adaptive optics imaging in the future. 

\begin{acknowledgements}
This research is based in part on data collected at the Subaru Telescope, which is operated by the National Astronomical Observatory of Japan. This research has been supported in part by the World Premier International Research Center Initiative, MEXT, Japan. This research has made use of the SIMBAD database and Vizier service, operated at CDS, Strasbourg, France. The authors wish to recognize and acknowledge the very significant cultural role and reverence that the summit of Mauna Kea has always had within the indigenous Hawaiian community. We are most fortunate to have the opportunity to conduct observations from this mountain. We are grateful to Collette Salyk, Glenn Schneider and Barb Whitney for their insightful comments. KBF and LMC were supported through the NASA Origins of Solar Systems program NNX09AB31G, and KBF through an NSF EAPSI Fellowship. CAG has been supported by NSF-AST 1008440 and through the NASA Origins of Solar Systems program on NNG13PB64P. MT is supported from Ministry of Science and Technology
(MoST) of Taiwan (Grant No. 103-2112-M-001-029). MWMcE is supported under NASA RTOP 12-OSS12-0045 through the NASA Origins of Solar Systems program. JC was supported by the National Science Foundation under Award No. 1009203. JPW is supported under NSF AST 1009314 and the NASA Origins of Solar System program under NNX13AK17G. MLS was supported under NASA ADP grant NNX09AC73G.
\end{acknowledgements}





\appendix
\section{10$\mu$m Feature}
\label{10micron}
As described in the main body of the text, we were able to create a model that fit both the global SED and the scattered light imagery well except in the region of the 10$\mu$m feature, which is clearly being overproduced. This section describes our efforts to suppress the feature in the SED.

Model imagery in a 10$\mu$m narrowband filter reveals the gap wall as the dominant source of 10$\mu$m emission in our models. The location of this wall is well constrained by our images, and we find that other tunable parameters affect the strength of the 10$\mu$m feature only minimally (scale height, flaring, etc.). The simplest way to affect the strength of the feature seems to be modification of the grain prescription, as we were unable to suppress the 10$\mu$m feature to the observed level with a standard silicate-rich small grain dust prescription \citep[e.g.][]{Kim:1994}. 

\begin{figure*}
\begin{center}
\includegraphics[scale=1.0]{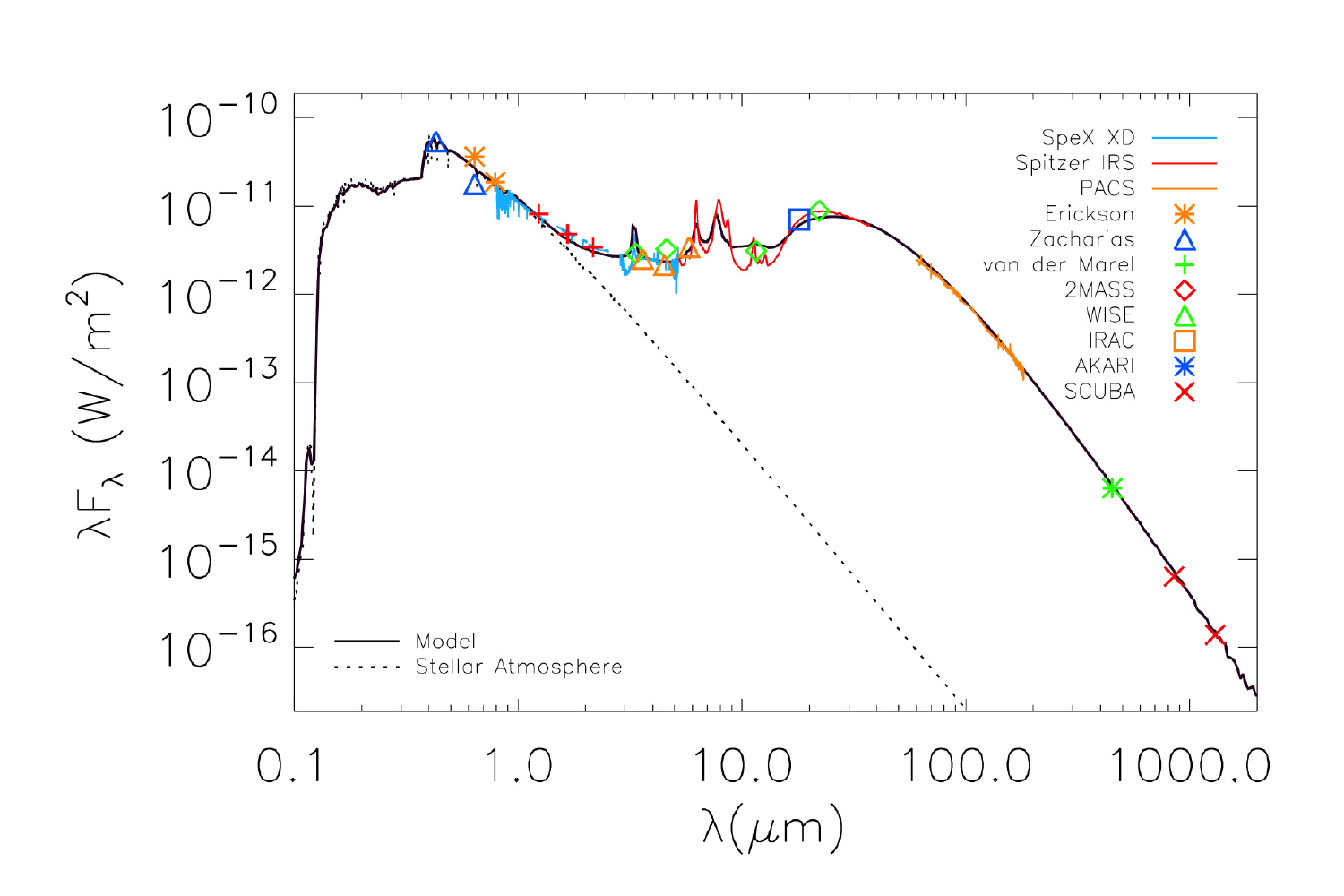}
\end{center}
\caption{Best-fitting Whitney model SED for \citet{Wood:2002} Model 1 grains. This provides a better fit at 10$\mu$m, but the IRS spectrum is poorly matched and the corresponding model imagery, shown in Figure \ref{fig:modelims2}, is a much poorer fit to our observations.}
\label{fig:sed2}
\end{figure*}

Among the small grain dust prescriptions that are standard in the Whitney code, we were best able to reproduce the small 10$\mu$m feature with the prescription of \citet{Wood:2002}'s Model 1, a mixture of $\sim$55$\%$ amorphous carbon and $\sim$45$\%$ astronomical silicates with a power law exponent of 3.0, a maximum grain size of 1mm and an exponential cutoff with a turnover at 50$\mu$m. The best fitting SED generated with this small grain dust assumption, for which all parameters except for the small grain dust prescription and the cavity depletion factor (increased to 0.05) are the same as in Table \ref{table:modelpars}, is shown in Figure \ref{fig:sed2}. This model provides a better fit in the 10$\mu$m region, though whether it is a better overall fit is unclear, as the fit to the Spitzer IRS spectrum is poorer.

\begin{figure*}
\centering
\subfloat[][]{
\label{fig:modelims2-a}
\includegraphics[scale=0.45]{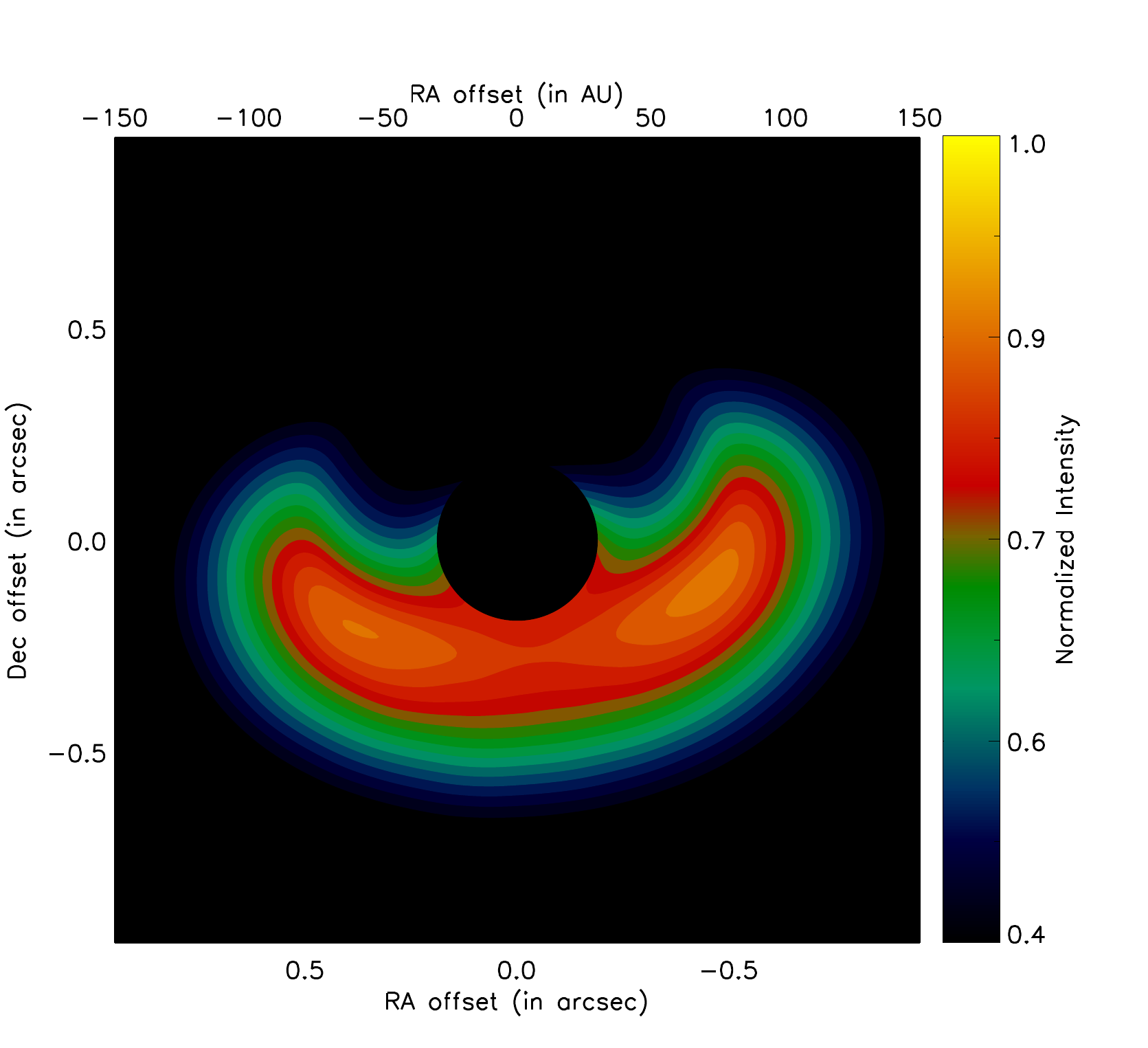}}
\subfloat[][]{
\label{fig:modelims2-b}
\includegraphics[scale=0.45]{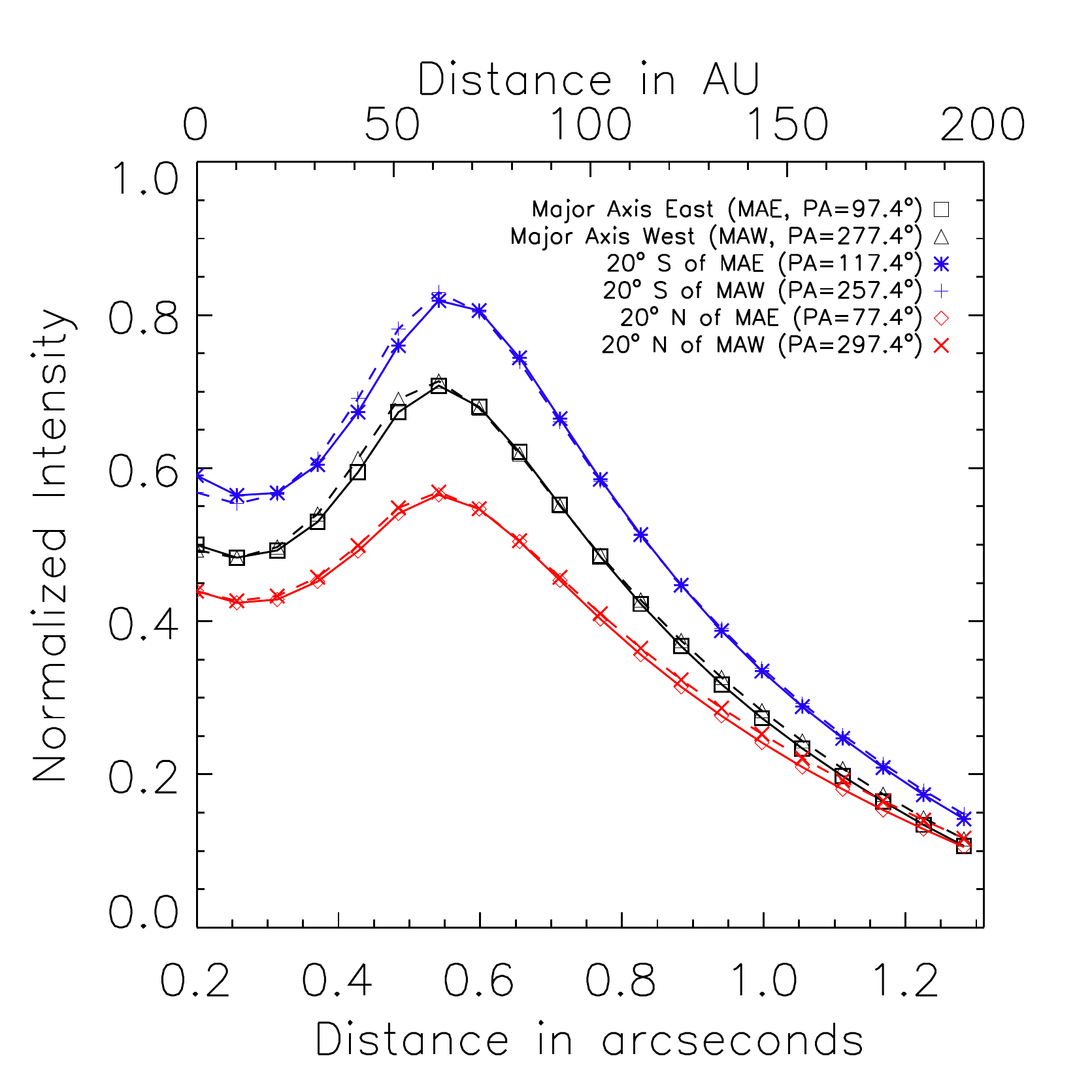}}\\
\caption[]{K-band model image for \citet{Wood:2002}'s Model 1 grains convolved with our PSF (left) and radial profiles taken through the same image (right). The H-band model and profiles are qualitatively similar. This model, though a better fit to the SED, is too bright in the South relative to the North, as revealed by the relative brightness of the model profiles along the major axis and to the south and north.}
\label{fig:modelims2}
\end{figure*}

The scattered light models generated under these assumptions, however, are a far poorer fit to our observed imagery. As shown in Figure \ref{fig:modelims2}, the disk becomes significantly brighter to the south and fainter to the north in these models than is observed. This may be partially due to the necessity of increasing the depletion factor inside the gap, which increases the shadowing of the exposed Northern cavity rim, or it may be that these grains are more strongly forward scattering. As the model imagery was clearly a poor fit, we did not investigate in detail. 

We thus reverted to the standard ISM-like dust prescription of \citet{Kim:1994}. We were able to mitigate the strength of 10$\mu$m emission somewhat by modifying it to grains from only 0.1-1.0$\mu$m in size. Our best fit models still overproduce 10$\mu$m emission, but the best fit model presented in the body of this paper is a much better fit to the observations as a whole. 

Oph IRS 48 is neither the only disk with this problem (see, for example \citet{Follette:2013}), nor is the fit in this region of heavy PAH emission particularly good in any existing disk models. None of the other SED models in the literature \citep[e.g.][]{Bruderer:2014, Maaskant:2013} include PAH emission, and since PAHs also emit strongly in the 10$\mu$m region, the 10$\mu$m excess in other models, though present, is not as apparent. 

We leave discussion of the mysterious lack of 10$\mu$m silicate emission in this and other disks for future work, though we note that one way to reduce it may be to increase the minimum grain size to $>$0.1$\mu$m. 

\appendix
\section{Radial Profiles}
\label{radprofs}

\begin{figure*}[t]
\centering
\subfloat[][]{
\label{fig:radprofs-a}
\includegraphics[scale=0.35]{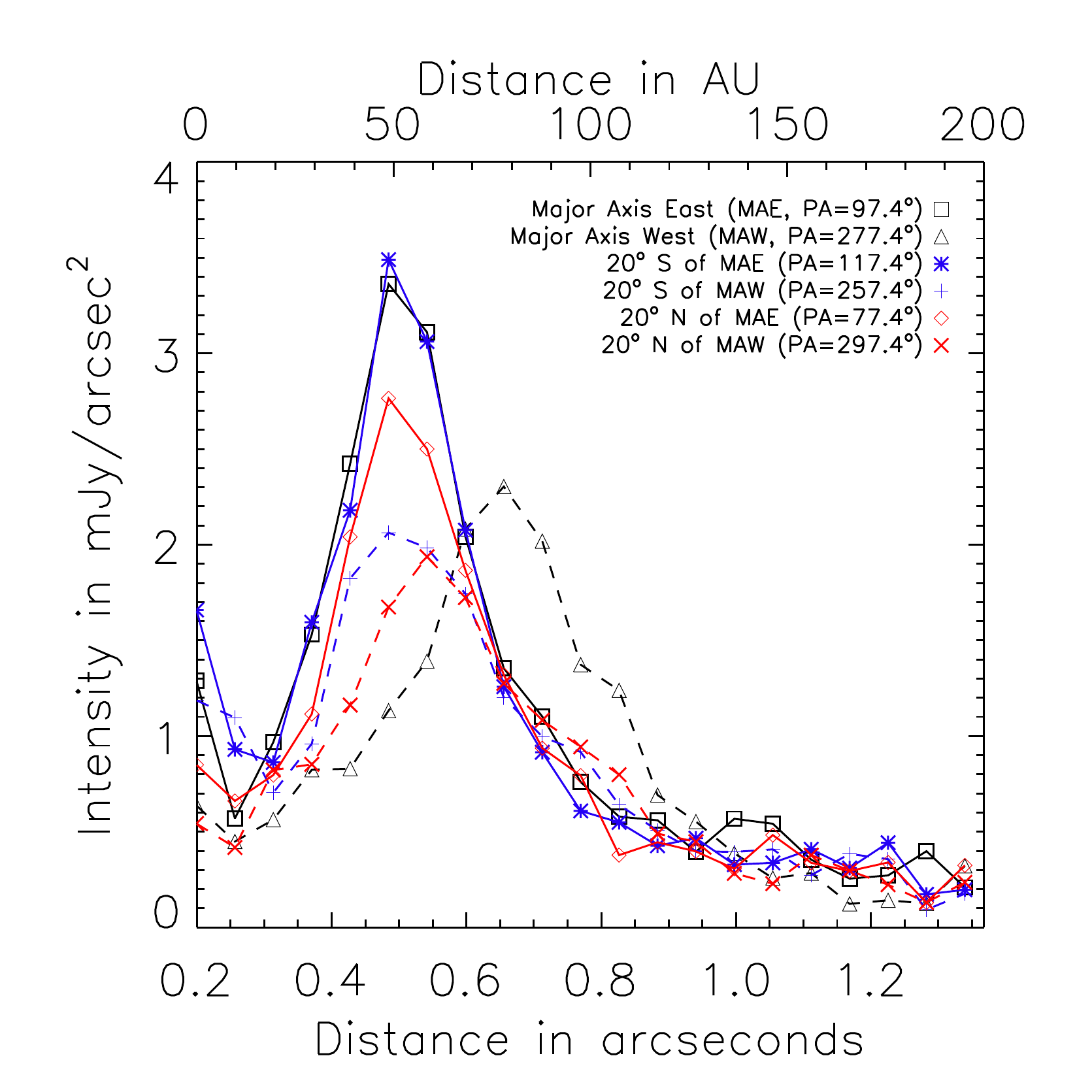}}
\subfloat[][]{
\label{fig:radprofs-b}
\includegraphics[scale=0.35]{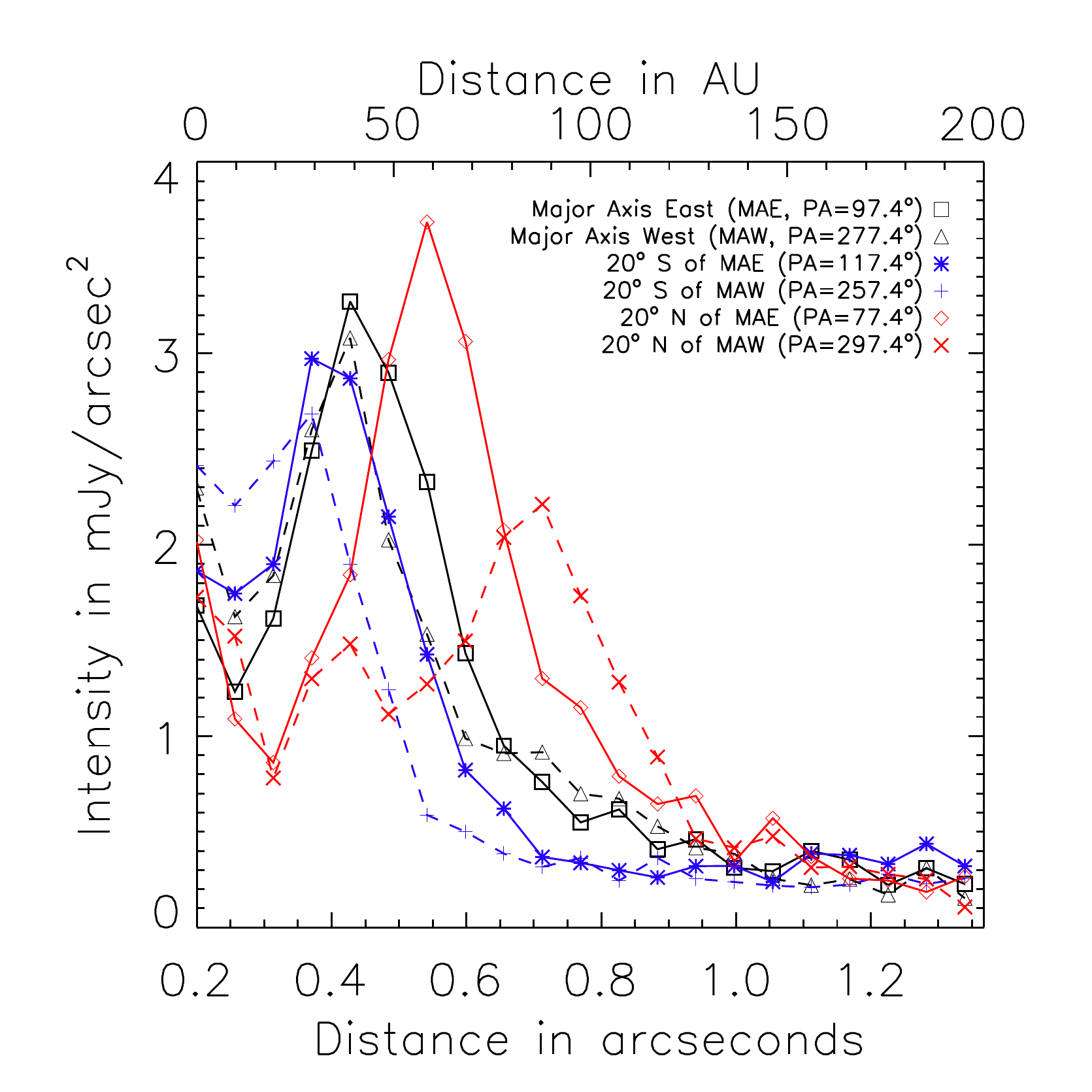}}\\
\subfloat[][]{
\label{fig:radprofs-c}
\includegraphics[scale=0.35]{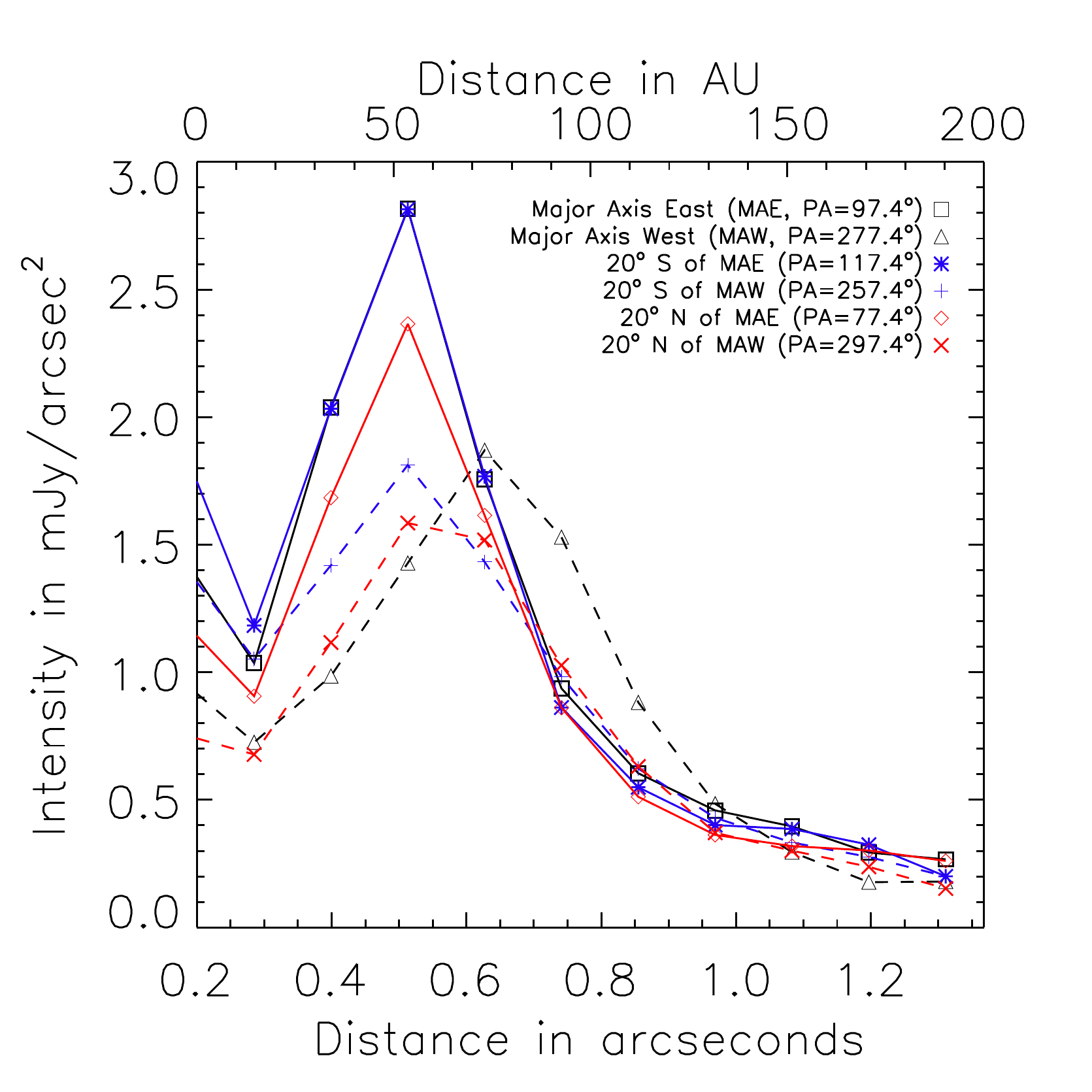}}
\subfloat[][]{
\label{fig:radprofs-d}
\includegraphics[scale=0.35]{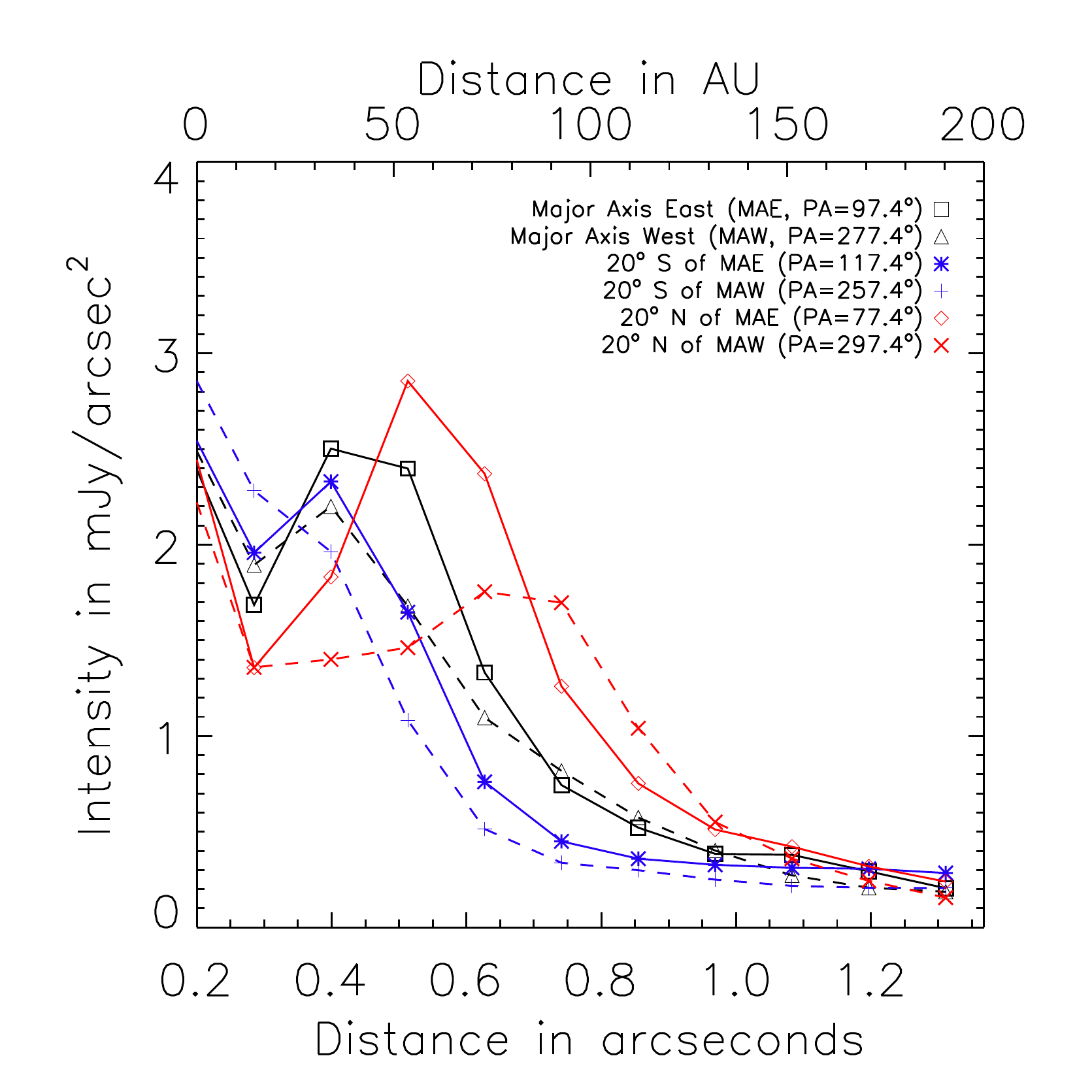}}\\
\subfloat[][]{
\label{fig:radprofs-e}
\includegraphics[scale=0.35]{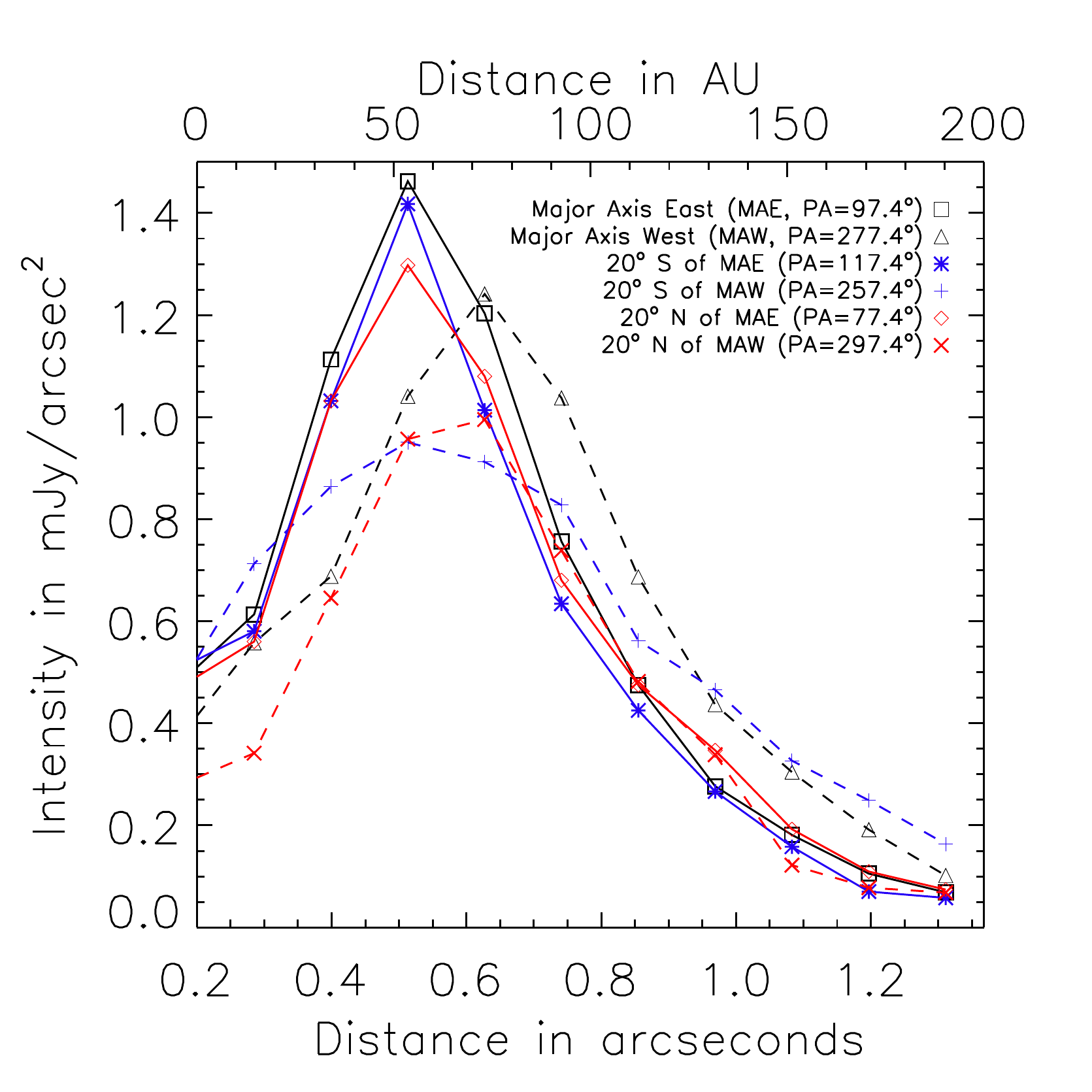}}
\subfloat[][]{
\label{fig:radprofs-f}
\includegraphics[scale=0.35]{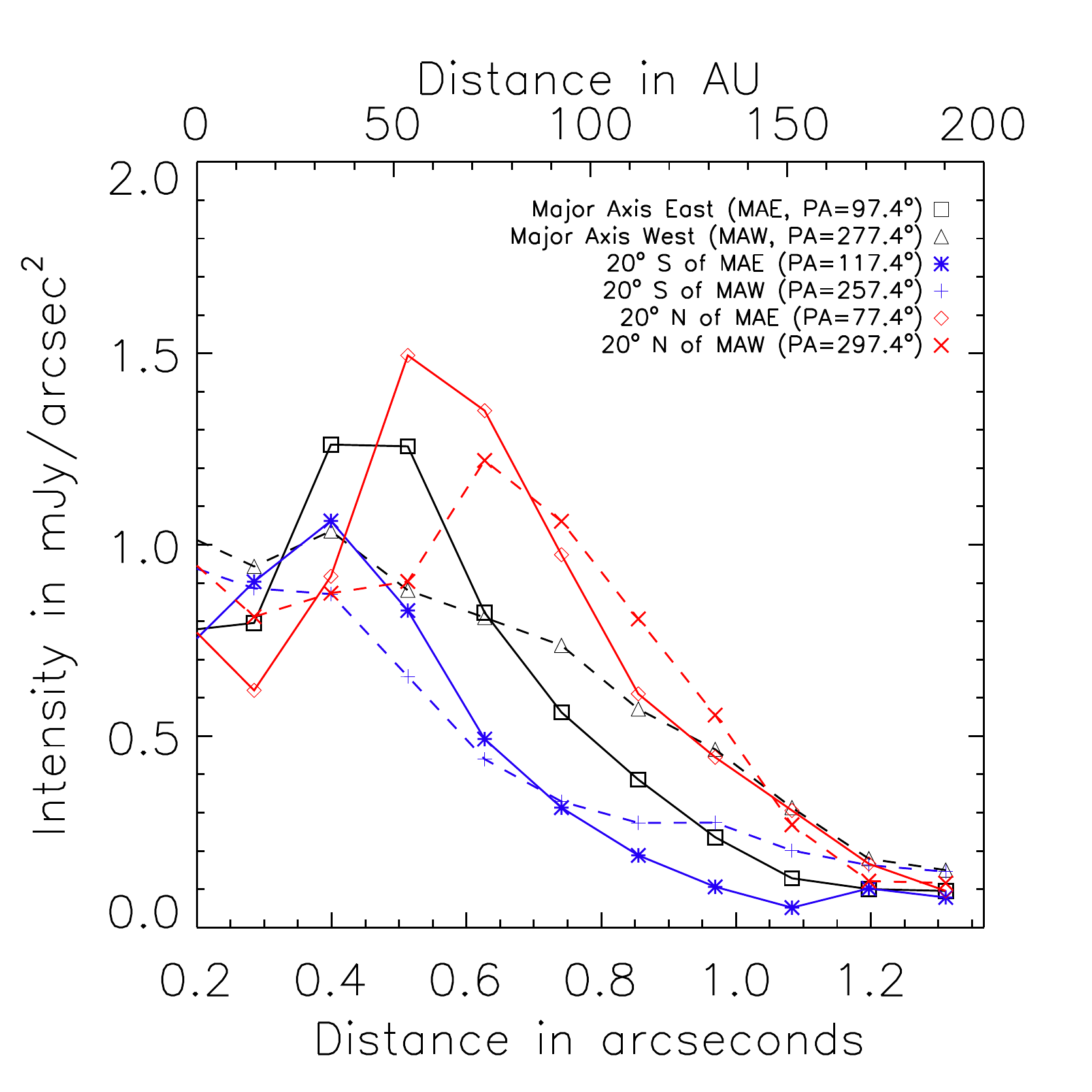}}
\label{fig:radprofs}
\caption[parbox=none]{Radial profiles taken through disk images deprojected to account for an inclination of 50$^{\circ}$ along PA=97.4$^{\circ}$. Profiles were taken both with respect to the center of the circular K-band cavity (left column), and  from the stellar location (right column). In all cases, solid lines represent profiles through the Eastern half of the disk and dashed lines represent profiles through the Western half of the disk. Black profiles were taken along the disk major axis and blue and red profiles were taken 20$^{\circ}$ North and South of the major axis, respectively.
\subref{fig:radprofs-a} and \subref{fig:radprofs-b} Radial profiles through the Ks-band halo-subtracted and deprojected PI image. Profiles have been binned to 6 pixels (0$\farcs$06) radially, equivalent to half of the stellar FWHM.;
\subref{fig:radprofs-c} and \subref{fig:radprofs-d} Radial profiles through a smoothed Ks-band halo-subtracted and deprojected PI image. The original image was smoothed with a 5.65 pixel (0$\farcs$05) Gaussian in order to match the stellar FWHM of the H-band image and allow for a direct comparison of the Ks and H-band profiles. Profiles have been binned to 12 pixels/0$\farcs$11 to match the H-band data.;
\subref{fig:radprofs-e} and \subref{fig:radprofs-f} Radial profiles through the H-band halo-subtracted and deprojected PI image. Profiles have been binned to 12 pixels/0$\farcs$11 radially, equivalent to half of the stellar FWHM.}
\end{figure*}

The radial profiles described in Section \ref{rps} were investigated relative to both the stellar residual and the center of the best fit to the cavity. Between these two potential reference points, profiles using the center of the cavity as the origin allow for clearer interprofile comparison, as they all peak at approximately the same radial location. Profiles taken relative to the cavity center also more closely match the circularly symmetric scattered light models discussed in Section \ref{diskmodel}. For completeness, we show profiles with respect to both reference points in figure \ref{fig:radprofs}. In the text, we have elected to use the lefthand (cavity referenced) profiles in our analysis. 

In all cases, radial profiles are shown along the major axis to both the East and West of the origin (either the stellar residual or the center of the cavity), and 20 degrees offset from the major axis to both the North and South in each case. 

Profiles taken through the final disk images at the two wavelengths also do not allow for a direct comparison between the structures at H and Ks bands and are potentially misleading. As the K band data are a factor of 2 smaller in FWHM, the rim appears much sharper in the profiles shown in Figures \ref{fig:radprofs-a} and \ref{fig:radprofs-b} than the H-band data in Figures \ref{fig:radprofs-e} and \ref{fig:radprofs-f}, so peak intensities cannot be directly compared. 

In order to construct a direct comparison, we convolved our K band image with a Gaussian to simulate image quality degradation. The FWHM of the Gaussian (5.65pixels/0$\farcs$05) was chosen in order to match the FWHM of the median combined intensity image in both image sets, and these convolved profiles are shown in Figures \ref{fig:radprofs-c} and \ref{fig:radprofs-d}. 

Convolving the Ks-band PI images with this Gaussian results in radial profiles that are remarkably similar between H and Ks bands for the Eastern half of the disk, and it is these two sets of radial profiles (Figure \ref{fig:radprofs-c} and \ref{fig:radprofs-e} that we have elected to use in the main body of the text.

A comparison between the convolved and unconvolved profiles also elucidates the reason why the Southern arc is not visible in the H-band data. Smoothing the K-band data in this way causes the arc to bleed into the stellar residual, erasing any trace of the cavity in this region. It is thus unsurprising that we did not resolve the Southern arc in the H-band data.

\bibliographystyle{apj}
\bibliography{IRS48} 

\end{document}